\documentclass[aps,prx,twocolumn,showpacs]{revtex4-1}
\pdfoutput=1
\usepackage{hyperref}
\hypersetup{
colorlinks = true, linkcolor = [rgb]{0,0,1}, citecolor = [rgb]{0,0,1},
                            urlcolor = [rgb]{0,0,1}}
\usepackage{graphicx,epsf}
\usepackage{amsmath}
\usepackage{amssymb}
\usepackage{color}
\usepackage{esint}
\usepackage{wasysym}

\newcommand{\bS}{{\bf S}}
\newcommand{\bR}{{\bf R}}
\newcommand{\bq}{{\bf q}}
\newcommand{\bQ}{{\bf Q}}
\newcommand{\bp}{{\bf p}}
\newcommand{\bk}{{\bf k}}

\newcommand{\taub}{\mbox{\boldmath $\tau $}}

\begin{document}

\title{Mean-field theory of interacting triplons in a
         two-dimensional valence-bond solid: 
         stability and properties of many-triplon states}

\author{R. L. Doretto}
\affiliation{Instituto de F\'isica Gleb Wataghin,
                  Universidade Estadual de Campinas,
                  13083-859 Campinas, SP, Brazil}

\date{\today}

\begin{abstract}
We study a system of $\bar{N}$ interacting triplons (the elementary excitations
of a valence-bond solid) described by an effective interacting boson model derived
within the bond-operator formalism in order to determine the stability
and the properties of many-triplon states.    
In particular, we consider the square lattice spin-$1/2$ $J_1$-$J_2$
antiferromagnetic Heisenberg model, focus on the intermediate
parameter region, where a quantum paramagnetic phase sets
in, and consider the columnar valence-bond solid as a reference state.
Within the bond-operator theory, the Heisenberg model is mapped into
an effective boson model in terms of triplet operators $t$. 
The effective boson model is studied at the harmonic
approximation and the energy of the triplons and the expansion of
the triplon operators $b$ in terms of the triplet operators $t$ are determined. 
Such an expansion allows us to performed a second mapping, and
therefore, determine an effective interacting boson model in terms of the triplon
operators $b$.   
We then consider systems with a fixed number $\bar{N}$ of triplons and  
study the stability of many-triplon states within a mean-field
approximation. We show that  
many-triplon states are stable, the lowest-energy ones are
constituted by a small number of triplons, and the excitation gaps are finite. 
For $J_2 = 0.48 J_1$ and $J_2 = 0.52 J_1$, we also calculate spin-spin
and dimer-dimer correlation functions, dimer order parameters, 
and the bipartite von Neumann entanglement entropy within 
our mean-field formalism in order to determine the properties
of the many-triplon state as a function of the triplon number $\bar{N}$.
We find that the spin and the dimer correlations decay exponentially
and that the entanglement entropy obeys an area law, regardless the
triplon number $\bar{N}$. Moreover, only for $J_2 = 0.48 J_1$, the spin correlations indicate
that the many-triplon states with large triplon number $\bar{N}$
might display a more homogeneous singlet pattern than the 
columnar valence-bond solid.
We also comment on possible relations between the many-triplon 
states with large triplon number $\bar{N}$ and gapped spin-liquid states.

\end{abstract}

\maketitle

\section{Introduction}
\label{sec:intro}

\begin{figure*}[t]
\centerline{\includegraphics[width=5.0cm]{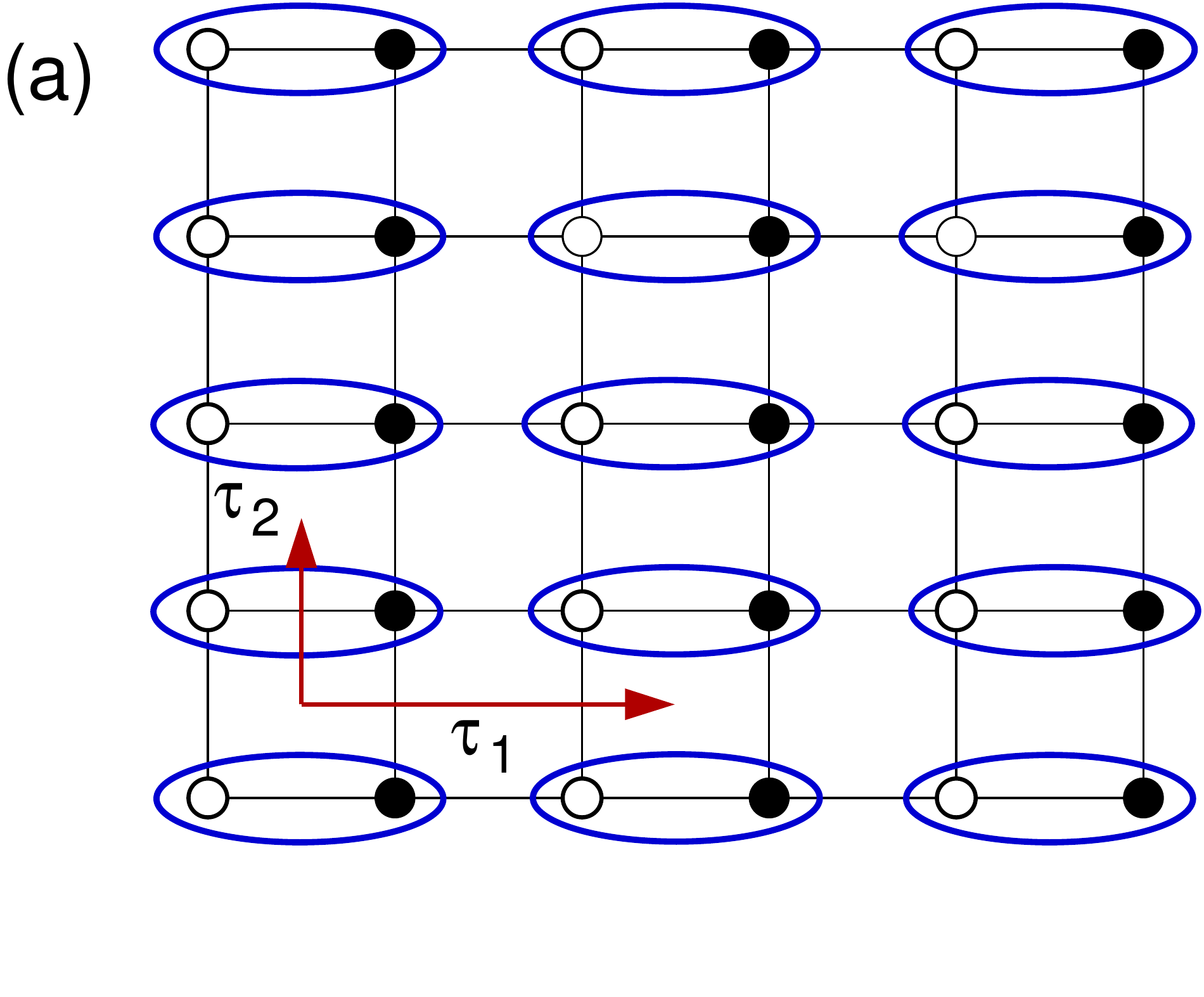}
            \hskip1.5cm
            \includegraphics[width=5.7cm]{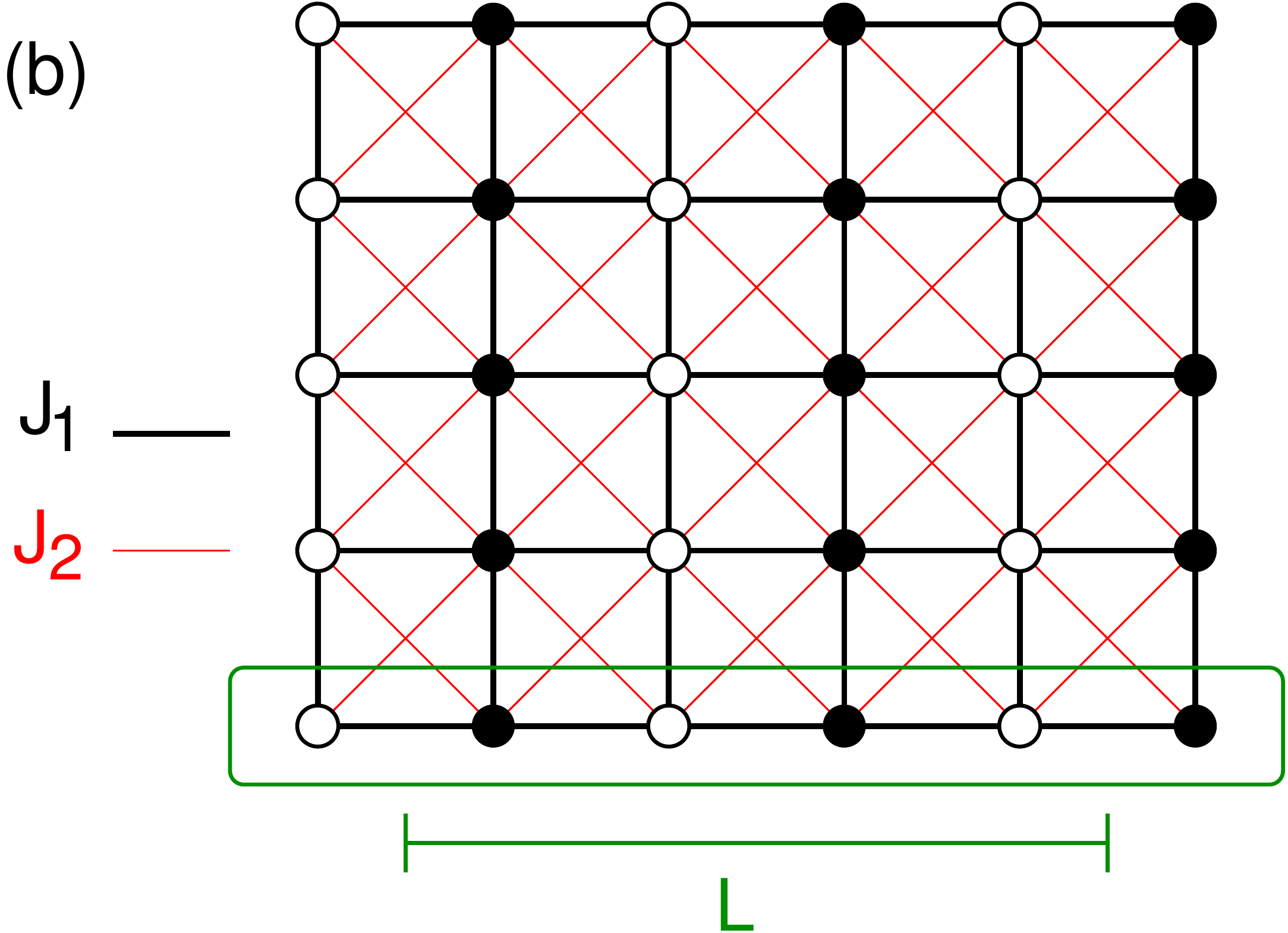}
            \hskip1.5cm
            \includegraphics[width=3.5cm]{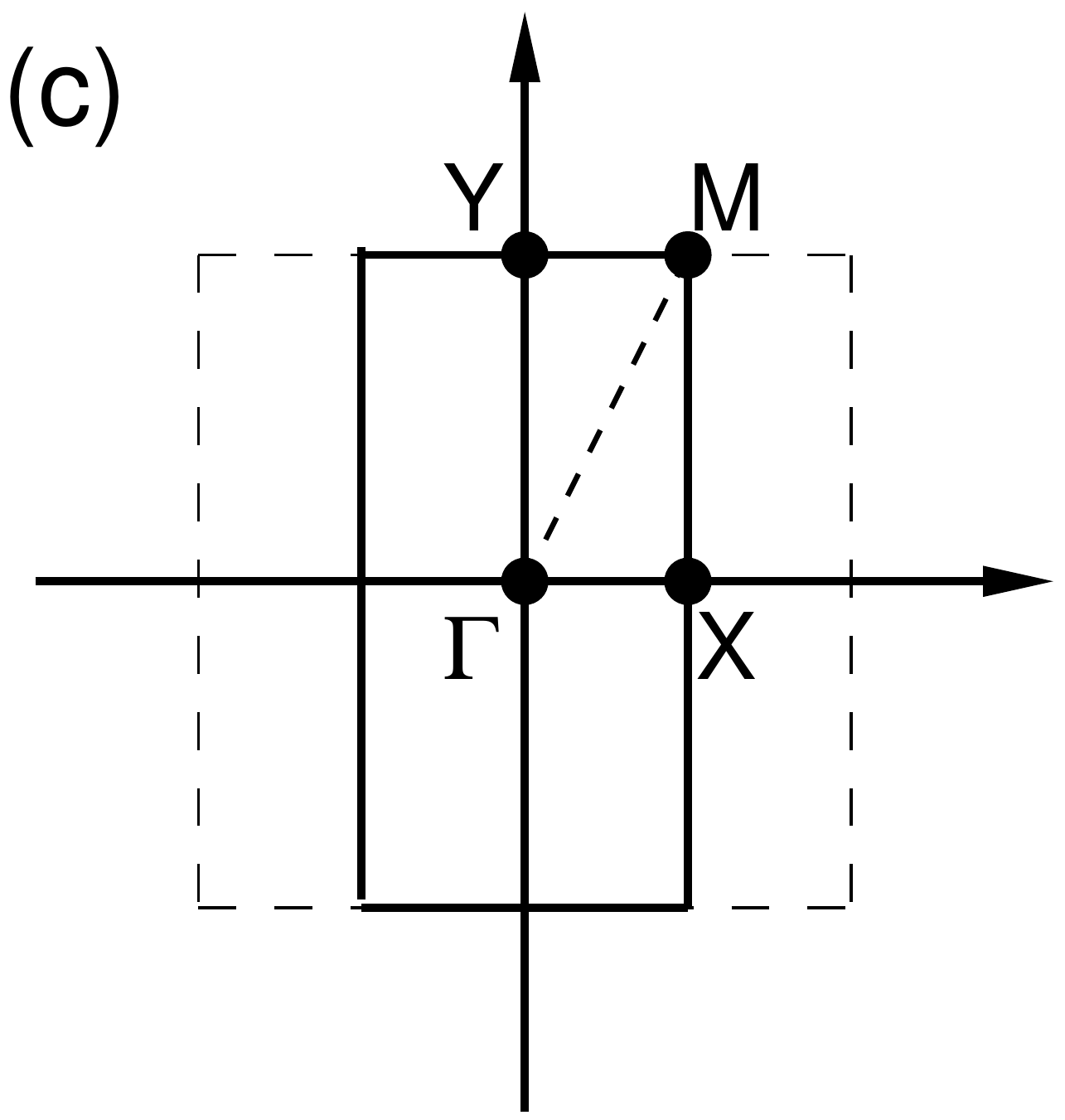}
}
\caption{Schematic representations of   
                (a) the columnar VBS ground state and 
                (b) the square lattice spin-$1/2$ $J_1$--$J_2$ AFM Heisenberg model \eqref{ham-j1j2}.
                The blue ellipses indicate that the spins $\bS^1$ (open circle) and 
                $\bS^2$ (filled circle) form a singlet state and
                $\taub_1$ and $\taub_2$ are the primitive vectors of
                the dimerized lattice $\mathcal{D}$ defined by the
                (blue) singlets.
                The green rectangle indicates the one-dimensional
                subsystem $A$ (dimer chain of size $L$) considered in
                the calculations of the entanglement entropy.
                (c) Brillouin zones of the dimerized (solid line) and the
                original square (dashed line) lattices. Here $\mathbf{X} = (\pi/2,0)$, $\mathbf{M} = (\pi/2,\pi)$,
                and $\mathbf{Y} = (0,\pi)$. The lattice spacing $a$ of the original square lattice is set to 1.
}
\label{fig:model}
\end{figure*}

A valence-bond solid (VBS) is a quantum paramagnetic (disordered)
phase that can be realized in a quantum spin system,
characterized by the absence of magnetic long-range order, 
but broken lattice symmetries \cite{review-sachdev}.
Such a state can be viewed as a regular arrangement of singlets that are
formed by a set of neighbor spins in a given lattice. 
An interesting example is the columnar VBS state on a square lattice
illustrated in Fig.~\ref{fig:model}(a): here, nearest-neighbor 
$S = 1/2$ spins are combined into a singlet (dimer) state, the unit
cell has two sites, and both translational and rotational lattice
symmetries are broken.

In two-dimensional quantum spin systems, VBS phases have been studied
since the seminal work of Read and Sachdev \cite{read89}.  
In particular, two-dimensional frustrated quantum antiferromagnets 
(AFMs) \cite{review-frustrated}  can, in principle,  host VBS phases,
since here the interplay between frustration
and quantum fluctuations could destroy magnetic long-range order.   
For instance, for the square lattice spin--$1/2$ $J_1$--$J_2$
AFM Heisenberg model, it was proposed that the ground state within the
intermediate parameter region $0.4\, J_1 \lesssim J_2 \lesssim 0.6\, J_1$ 
could be either a (dimerized) columnar VBS or a (tetramerized) plaquette VBS 
(see Sec.~\ref{sec:model} below for more details).
Interesting, for the same model but on the {\sl honeycomb} lattice, 
density matrix renormalization group (DMRG) calculations 
\cite{ganesh13,zhu13}  indicate that the ground state of the model is a
dimerized VBS for $J_2 \gtrsim 0.36\, J_1$ \cite{ferrari17}.    
A third example of a frustrated two-dimensional quantum magnet is the
spin-$1/2$ nearest-neighbor AFM Heisenberg model on the kagome lattice. 
Here, a dimerized VBS with a 36-site unit cell has been proposed as the
ground state \cite{marston91,yang08,hwang15}. 
Although it has been receiving a lot of attention in recent years
(see, e.g., Ref.~\cite{He17} and the references therein), 
the ground state of the AFM Heisenberg model on the kagome lattice is
still under debate: in addition to the dimerized VBS with a 36-site unit cell, gapped and
gapless spin-liquid states \cite{review-balents,rmp17} have also been
proposed.

In addition to the above frustrated two-dimensional AFMs, the
nonfrustrated $J$-$Q$ model also hosts a dimerized VBS phase
\cite{sandvik07,sandvik10,sandvik12,suwa16}. 
The $J$-$Q$ model describes $S=1/2$ spins on a square lattice
interacting via a nearest-neighbor AFM Heisenberg ($J$) term
and an additional four-spin ($Q$) term within each plaquette.
Since it is a nonfrustrated quantum spin system, such a model is 
free from the so-called sign-problem \cite{henelius00}, and therefore,
it can be studied within quantum Monte Carlo (QMC) simulations. 
It was found that a columnar VBS phase sets in for small exchange
coupling $J$, while a N\'eel magnetic long-range order phase is stable
for large $J$, with the N\'eel--VBS quantum phase transition (QPT) taking
place at $J_c = 0.0447\, Q$ \cite{sandvik10}.
Indeed, QMC simulations for the $J$-$Q$ model 
\cite{sandvik07,sandvik10,suwa16} indicate that the N\'eel--VBS 
is a continuous QPT, in agreement with the 
deconfined quantum criticality scenario \cite{dqc}: 
recall that, within the Landau-Ginzburg framework, 
the N\'eel--VBS should be a first-order  QPT, since the
N\'eel and the VBS phases are described by two distinct order
parameters.

The elementary excitations of a dimerized VBS phase
correspond to singlets turned into triplets, 
the so-called {\sl triplons} \cite{review-sachdev}. 
Such excitations can be analytically described, for instance,
within the bond-operator representation \cite{sachdev90}, 
where spin operators are expanded in terms of singlet ($s$) and 
triplet ($t$) boson operators. 
This formalism follows the ideas of the
Holstein--Primakoff representation for spins \cite{assa}, which describes
fluctuations (spin-waves) above a (semiclassical) ground state with magnetic
long-range order. 
The bond-operator representation, however, describes
fluctuations above a quantum paramagnetic ground state.  
For a VBS phase, the bond-operator formalism allows us to map a spin
Hamiltonian into an interacting Hamiltonian in terms of triplet
operators $t$, as exemplified below. 
From the lowest-order (quadratic) terms of the interacting triplet
Hamiltonian, we determine the {\sl triplon} (boson) operators $b$ in
terms of the triplet operators $t$ and find the triplon spectrum.  
The effects of the cubic and quartic triplet--triplet interactions can
be perturbatively taken into account (as done, e.g., in
Refs.~\cite{doretto12} and \cite{doretto14} respectively for a
dimerized and a tetramerized VBS phases), and therefore, corrections
to the (harmonic) energy of the triplons can be determined.  
In particular, the procedure employed in Ref.~\cite{doretto12} 
allows us to systematically determined an interacting boson model for the
triplons.

In this paper, we study the effective interacting boson model for the
triplons $b$ derived within the bond-operator formalism for a given 
VBS (reference) state but, instead of following the procedure employed in
Refs.~\cite{doretto12,doretto14},  we consider systems with a fixed
number $\bar{N}$ of {\sl triplons} $b$ and determine the stability and
the properties of possible  {\sl many-triplon states} within a
mean-field approximation.  
The first motivation for our study is to check whether a state with a
large number $\bar{N}$ of triplons $b$ could restore some of the
lattice symmetries broken when the VBS state sets in: Once 
a given number of triplons are excited above the VBS ground state, the
triplon-triplon interaction could yield two-triplon bound states 
with total spin zero;  the new ground state should also be formed by a
set of singlets, similar to the VBS state, but it should no longer
display the original VBS pattern; indeed, short (nearest-neighbor) and 
long singlets might be present, and therefore, some of the lattice 
symmetries might be restored.
The second motivation for our study is to verify whether such a
many-triplon state could describe a (gapped) spin-liquid phase: if so,
then the procedure discussed in this paper could be employed to study
spin-liquid phases within the bond operator formalism, i.e., it would be
an alternative to the Schwinger boson formalism \cite{assa}
that is used to analytically describe gapped spin-liquid phases \cite{yang16}.

In order to determine the stability of many-triplon states and
their possible relation with gapped spin-liquid phases,
we consider, in particular, the spin-$1/2$ $J_1$-$J_2$ AFM Heisenberg
model on the square lattice, since, in principle, it can host a VBS phase. 
Moreover, we concentrate on a {\sl columnar} VBS phase, which
is considered as the reference state.   
In addition to discuss the stability of possible many-triplon
states, we also determine its features in terms of the triplon
number $\bar{N}$: the corresponding 
spin-spin and dimer-dimer correlation functions, 
dimer order parameters, and 
the bipartite von-Neumann entanglement entropy
are determined within our mean-field formalism.

\subsection{Overview of the results}
\label{sec:overview}

Within the bond-operator formalism, we firstly map the square lattice
spin-$1/2$ AFM Heisenberg model [Eq.~\eqref{ham-j1j2}] into
an effective boson model in terms of triplet operators $t$
[Eq.~\eqref{h-effective} and Eqs.~\eqref{h0}-\eqref{h4}], which is studied
at the harmonic approximation, and then performe a second mapping that
yields an effective boson model in terms of {\sl triplon}
operators $b$ [Eq.~\eqref{h-effective2}], which is studied at a
mean-field approximation. Our main findings are the following: \\
(a) {\sl Harmonic approximation for the effective boson model I}: We
calculate the ground-state energy of the columnar VBS
[Fig.~\ref{fig:egs}(a)] in terms of $J_2/J_1$ and the energy of the
triplons (Fig.~\ref{fig:disp}). We find that 
the columnar VBS is stable for $0.30\, J_1 \le  J_2 \le  0.63\, J_1$
and that the excitation spectra are gapped (Fig.~\ref{fig:gap}).\\
(b) {\sl Mean-field approximation for the effective boson model II:}
We find that many-triplon states are stable as long as the triplon
number $\bar{N} \le \bar{N}_{MAX,2}$ (Fig.~\ref{fig:nbarmax}), 
the lowest-energy ones are constituted by a small number of triplons
[Figs.~\ref{fig:egs}(b) and \ref{fig:nbarmax}], and the
excitation spectra above the many-triplon states are gapped
(Figs.~\ref{fig:disp} and \ref{fig:gap}).\\
(c) {\sl Properties of the many-triplon states}: 
For $J_2 = 0.48 J_1$ and $J_2 = 0.52 J_1$, we find that the spin-spin
(Figs.~\ref{fig:cor-spin01} and \ref{fig:cor-spin02}) and the
dimer-dimer (Fig.~\ref{fig:cor-dimer}) correlation functions decay
exponentially and that the bipartite von Neumann entanglement entropies
(Fig.~\ref{fig:entropy}) obey an area law, regardless the triplon
number $\bar{N}$. Interesting, the spin-spin correlation function $C_x(r)$
[Figs.~\ref{fig:cor-spin01}(a) and \ref{fig:cor-spin02}(a)] and the
dimer-dimer correlation function $D_{xx}(r)$ (Fig.~\ref{fig:cor-dimer})
indicate that the many-triplon states with large triplon
number $\bar{N}$ might display a more homogeneous singlet pattern than
the columnar VBS only for $J_2 = 0.48 J_1$.
Our analysis indicates that, within the quantum paramagnet
(disordered) parameter region of 
the square lattice $J_1$-$J_2$ model, configurations with $J_2
\lesssim 0.51  J_1$ and $J_2 \gtrsim 0.51 J_1$ display distinct
features as recently found on numerical
calculations \cite{gong14, wang18, ferrari20,nomura20}. \\
(d) {\sl Possible relation with spin-liquid phases}: The results for
the dimer order parameters (Fig.~\ref{fig:orderpar})
and the bipartite von Neumann entanglement entropies
(Fig.~\ref{fig:entropy}) indicate that many-triplon states with
large triplon number $\bar{N}$ do not describe a gapped spin-liquid
phase for the square lattice $J_1$-$J_2$ model when a columnar VBS is
considered as a reference state.

The reader not interested in technical details may skip
Secs.~\ref{sec:bond}--\ref{sec:entropy} and go straight to
Sec.~\ref{sec:summary}.

\subsection{Outline}
\label{sec:outline}

Our paper is organized as follows: 
In Sec.~\ref{sec:bond}, we briefly summarize the bond-operator
representation \cite{sachdev90} for spin operators. 
A short review about the square lattice spin-$1/2$ $J_1$--$J_2$ AFM
Heisenberg model is presented in Sec.~\ref{sec:model}.
In Sec.~\ref{sec:boson-model}, an effective boson model in terms of 
the triplet operators $t$ for the columnar VBS phase is derived and it
is studied within the (lowest-order) harmonic approximation. Here,
we define the triplon operators $b$ in terms of the triplet
operators $t$ and determine the energy of the triplons. 
An effective interacting boson model for the triplon operators $b$
is derived in Sec.~\ref{sec:model-b}. 
We then consider systems with a fixed number $\bar{N}$ of triplons, 
discuss the stability of the many-triplon states, and determined the
excitation spectra within a mean-field approximation.
In Sec.~\ref{sec:corr}, spin-spin and dimer-dimer correlation
functions and dimer order parameters of the columnar VBS ground state and
the many-triplon state with different values of the triplon number
$\bar{N}$ are determined.  
Sec.~\ref{sec:entropy} is devoted to the calculation of the bipartite
von Neumann entanglement entropy of the columnar VBS ground state and the
many-triplon states. Here a one-dimensional (line) subsystem is
considered, a choice that allows us to analytically determined the entanglement
entropies.
We comment on possible implications of our results for the
$J_1$-$J_2$ model and provide a brief summary of our main
findings in Sec~\ref{sec:summary}. 
Some details of the results discussed in the main text are presented
in the two Appendices.

\section{Bond operator representation}
\label{sec:bond}

We start by briefly reviewing the bond-operator representation for
spins introduced by Sachdev and Bhatt \cite{sachdev90}. 
Our summary closely follows the lines of Ref.~\cite{leite19}.

Let us consider the Hilbert space of two $S=1/2$ spins, $\mathbf S^1$
and $\mathbf S^2$, which is made out of a singlet and three triplet states,
\begin{eqnarray}
| s \rangle   &=& \frac{1}{\sqrt{2}}\left(|\uparrow \downarrow \rangle
                   -|\downarrow \uparrow \rangle \right), 
\;\;\;\;\;\;\;
| t_x \rangle = \frac{1}{\sqrt{2}}\left(|\downarrow \downarrow \rangle
                   -|\uparrow \uparrow \rangle \right), 
\nonumber \\
| t_y \rangle &=& \frac{i}{\sqrt{2}}\left(|\uparrow \uparrow \rangle
                   +|\downarrow \downarrow \rangle \right), 
\;\;\;\;\;\;\;
| t_z \rangle = \frac{1}{\sqrt{2}}\left(|\uparrow \downarrow \rangle
                   +|\downarrow \uparrow \rangle \right).
\nonumber \\
\end{eqnarray}
One can define a set of boson operators,
$s^\dagger$ and $t^\dagger_\alpha$, with $\alpha = x$, $y$, $z$,
which respectively creates singlet and triplet states out of a fictitious vacuum
$|0\rangle$, i.e,
\begin{equation}
  | s \rangle = s^\dagger |0\rangle \;\;\;\;\;\; {\rm and} 
  \;\;\;\;\;\;
  | t_\alpha \rangle = t_\alpha^\dagger |0\rangle,
\end{equation}
with $\alpha = x$, $y$, $z$. 
In order to remove unphysical states from the enlarged Hilbert space, 
the constraint
\begin{equation}
s^\dagger s + \sum_\alpha t^\dagger_\alpha t_\alpha = 1
\label{constraint}
\end{equation}
should be introduced.   
Then, one calculates the matrix elements of each component of the two
spin operators within the basis $|s\rangle$ and $|t_\alpha\rangle$,
i.e., one determines $\langle s | S^\mu_\alpha | s \rangle$, 
$\langle s | S^\mu_\alpha | t_\beta \rangle$, and
$\langle t_\gamma | S^\mu_\alpha | t_\beta \rangle$,
with $\mu = 1$, $2$ and $\alpha$, $\beta$, $\gamma = $ $x$, $y$, $z$. 
The set of results allows us to conclude that 
the components of the spin operators $\mathbf S^1$ and $\mathbf S^2$
can be expressed in terms of the boson operators $s^\dagger$ and
$t^\dagger_\alpha$ as 
\begin{eqnarray}
S^{1,2}_\alpha &=& \pm\frac{1}{2}\left(s^\dagger t_\alpha + t^\dagger_\alpha s
          \mp i\epsilon_{\alpha\beta\gamma}t^\dagger_\beta t_\gamma \right),
\label{spin-bondop}
\end{eqnarray}
where $\epsilon_{\alpha\beta\gamma}$ is the completely antisymmetric
tensor with $\epsilon_{xyz} = 1$ and the summation convention over
repeated indices is considered. 
One then generalizes the bond-operator representation \eqref{spin-bondop} 
for the lattice case, and therefore, a spin Hamiltonian
can be easily written in terms of the boson operators  
$s^\dagger_i$ and $t^\dagger_{i\,\alpha}$.

\section{The $J_1$--$J_2$ square lattice antiferromagnet
            Heisenberg model}
\label{sec:model}

To study a system of interacting triplons, we 
consider, in particular, the spin--$1/2$ $J_1$--$J_2$ AFM
Heisenberg model on the square lattice, 
\begin{equation}
 \mathcal{H} = J_1\sum_{\langle ij \rangle} \bS_i\cdot\bS_j 
             + J_2\sum_{\langle\langle ij \rangle\rangle} \bS_i\cdot\bS_j,
\label{ham-j1j2}
\end{equation}
where $\bS_i$ is an spin--$1/2$ operator at site $i$ and $J_1 > 0$ and
$J_2 > 0$ are, respectively, the nearest--neighbor and
next--nearest--neighbor exchange couplings, see
Fig.~\ref{fig:model}(b).

It is well known that \cite{doretto14, wang16,yu16,yu19,choo19,trebst19,
kotov99,sheng18,eggert14, zhito96, gong14, ralko09,richter15,wang18,ferrari18,yuan18,yang16,ferrari20,nomura20},
at temperature $T=0$, the model \eqref{ham-j1j2} has a
semiclassical N\'eel  magnetic long-range ordered (LRO) phase with ordering
wave vector  $\bQ = (\pi,\pi)$ for  $J_2 \lesssim 0.4\, J_1$, 
a collinear magnetic LRO phase with $\bQ = (\pi,0)$ 
or $(0,\pi)$ for $J_2 \gtrsim 0.6\, J_1$, 
and a quantum paramagnetic phase 
for $0.4\, J_1 \lesssim J_2 \lesssim 0.6\, J_1$. 
The nature of the quantum paramagnetic phase is still under debate.
Indeed, several proposals have been made for the ground state of the model
\eqref{ham-j1j2} within this intermediate parameter region:   
a (dimerized) columnar VBS [Fig.~\ref{fig:model}(a)], where both translational and 
rotational lattice symmetries are broken \cite{kotov99,sheng18},
a (dimerized) staggered VBS \cite{eggert14},
a (tetramerized) plaquette VBS, where only the translational lattice
symmetry is broken \cite{doretto14, zhito96, gong14}, 
a mixed columnar-plaquette VBS \cite{ralko09},
and gapless \cite{richter15,wang18,ferrari18,yuan18} 
and gapped \cite{yang16} spin-liquid ground states.
Moreover, while there are indications that the quantum
paramagnetic-collinear is a first-order QPT,
it is not clear whether the N\'eel-quantum paramagnetic QPT
is a first-order or a continuous transition \cite{comment01}.

In the following, we concentrate on the intermediate parameter region 
$0.4\, J_1 \lesssim J_2 \lesssim 0.6\, J_1$
and, in particular, consider the columnar VBS phase 
[Fig.~\ref{fig:model}(a)].

\section{Effective boson model I}
\label{sec:boson-model}

In this section, we consider the bond-operator representation
\eqref{spin-bondop} and derive an effective boson Hamiltonian in terms of
the triplet operators $t_{i\,\alpha}$ to describe the columnar
VBS phase of the Heisenberg model \eqref{ham-j1j2}.

We start rewriting the Hamiltonian \eqref{ham-j1j2} in terms
of the underline dimerized lattice $\mathcal{D}$ defined by the singlets (dimers)
as shown in Fig.~\ref{fig:model}(a),   
\begin{align}
 \mathcal{H}   =  \sum_{i \in \mathcal{D}} 
     & J_1 \left(   \bS^1_i\cdot\bS^2_i  +  \bS^1_i\cdot\bS^1_{i+2}  +
                      \bS^2_i\cdot\bS^2_{i+2}  + \bS^2_i\cdot\bS^1_{i+1}  \right)
\nonumber \\
   &+  J_2\left(  \bS^1_i\cdot\bS^2_{i+2}  + \bS^2_i\cdot\bS^1_{i+2} \right)
\nonumber \\
   &+ J_2 \left( \bS^2_i\cdot\bS^1_{i+1+2}  + \bS^2_i\cdot\bS^1_{i+1-2} \right).
\label{ham-dimer}
\end{align}
Here, $i$ is a site of the dimerized lattice $\mathcal{D}$,
which has two spins per unit cell ($\bS^1_i$ and $\bS^2_i$),  
and the index $n = 1,2$ corresponds to the dimer 
nearest-neighbor vectors $\taub_n$, 
\begin{equation}
   \taub_1 = 2 a \hat{x},  \;\;\;\;\;\;\;\;\;\;\;
   \taub_2 = a \hat{y}, 			
\label{tau-col}
\end{equation} 	
with $a$ being the lattice spacing of the {\sl original}  square
lattice. Hereafter, we set $a=1$.

An effective model in terms of the singlet $s_i$ and triplet $t_{i \alpha}$ 
boson operators can be obtained by substituting the bond-operator
representation \eqref{spin-bondop} generalized to the lattice case into the
Hamiltonian \eqref{ham-dimer}. It is easy to show that the Hamiltonian \eqref{ham-dimer}
assumes the form
\begin{equation}
 \mathcal{H} = \mathcal{H}_0 + \mathcal{H}_2 + \mathcal{H}_3 + \mathcal{H}_4, 
\label{h-effective}
\end{equation}
where the $\mathcal{H}_n$ terms contain $n$ triplet operators 
[for details, see Eq.~\eqref{Heff-sites}].  
Moreover, we consider the constraint \eqref{constraint} on
average via a Lagrange multiplier $\mu$, i.e., we add the following
term to the Hamiltonian \eqref{h-effective}
\[
   -\mu \sum_i \left( s_i^\dagger s_i + t_{i \alpha}^\dagger t_{i \alpha}  - 1 \right).
\]
Within the bond-operator formalism, the columnar VBS ground state 
[Fig.~\ref{fig:model}(a)] can be viewed as a condensate of the
singlets $s_i$. Therefore, one sets 
\begin{equation}  
     s_i^\dagger = s_i  = \langle s_i^\dagger \rangle =  
    \langle s_i \rangle \rightarrow  \sqrt{N_0}  
\label{condensate}
\end{equation}
in the Hamiltonian \eqref{h-effective} and ends up with an
effective boson Hamiltonian only in terms of the triplet 
boson operators $t_{i\alpha}$. 
As discussed below, the constants $N_0$ and $\mu$ are
self-consistently determined for a fixed value of the ratio $J_2/J_1$
of the exchange couplings.

\begin{figure}[t]
\centerline{\includegraphics[width=8.5cm]{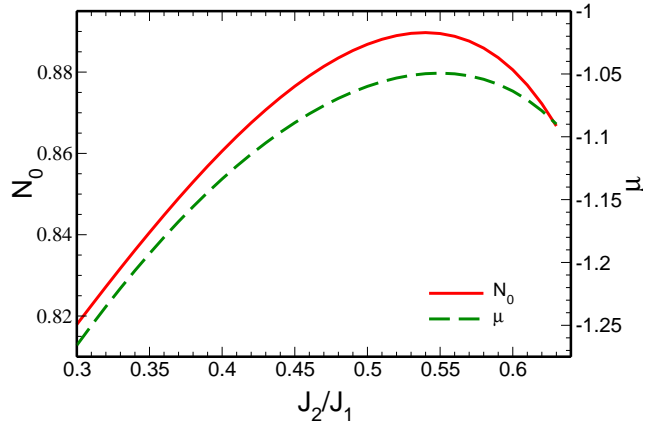}}
\caption{
  Parameters $N_0$ and $\mu$ as a functions of $J_2/J_1$ for the
  columnar VBS ground state (harmonic approximation) determined from
  the numerical solutions of the self-consistent equations
  \eqref{self-mu-nzero}.   
}
\label{fig:munzero}
\end{figure}

\begin{figure*}[t]
\centerline{\includegraphics[width=8.5cm]{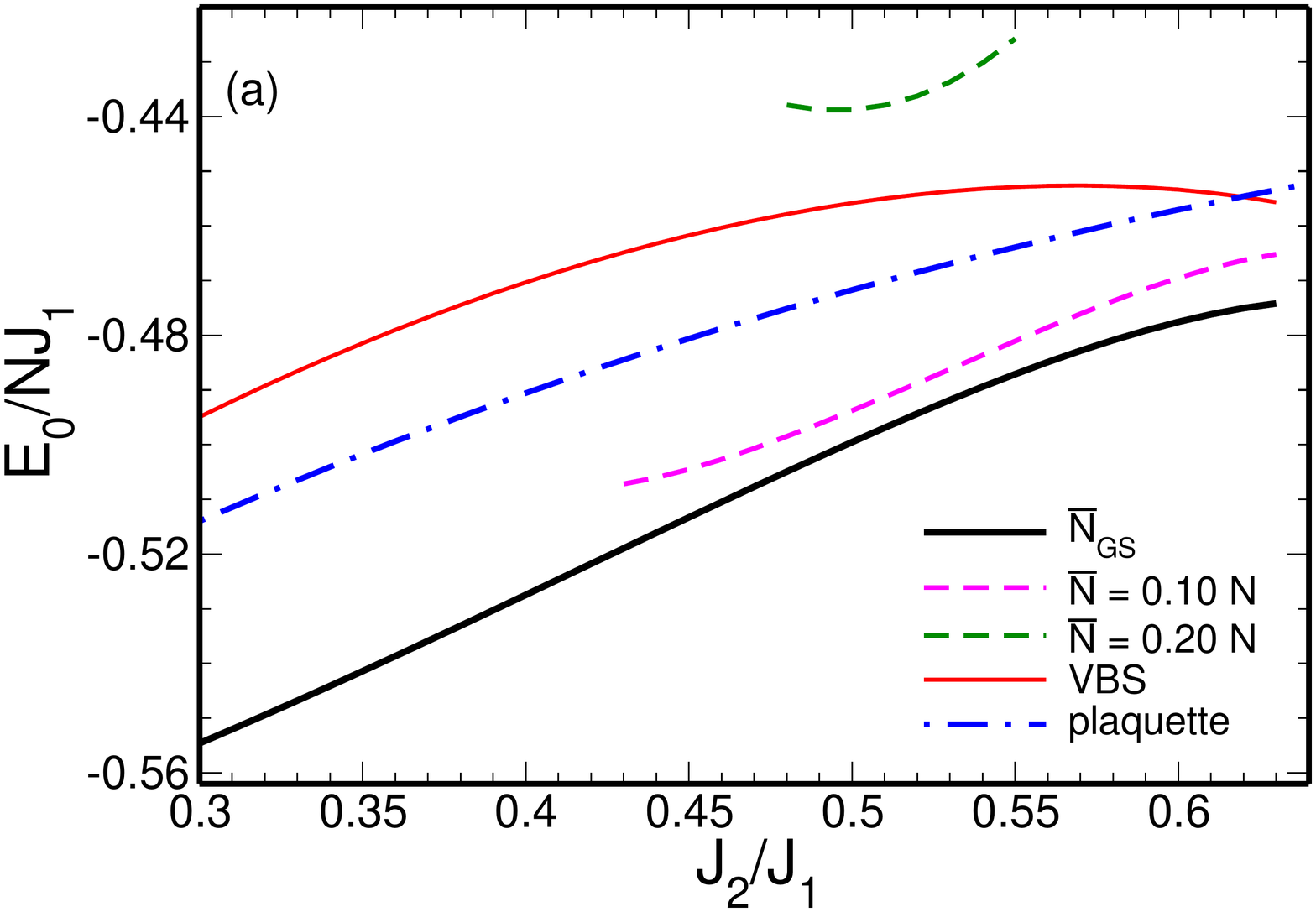}
                   \hskip0.5cm
                  \includegraphics[width=8.5cm]{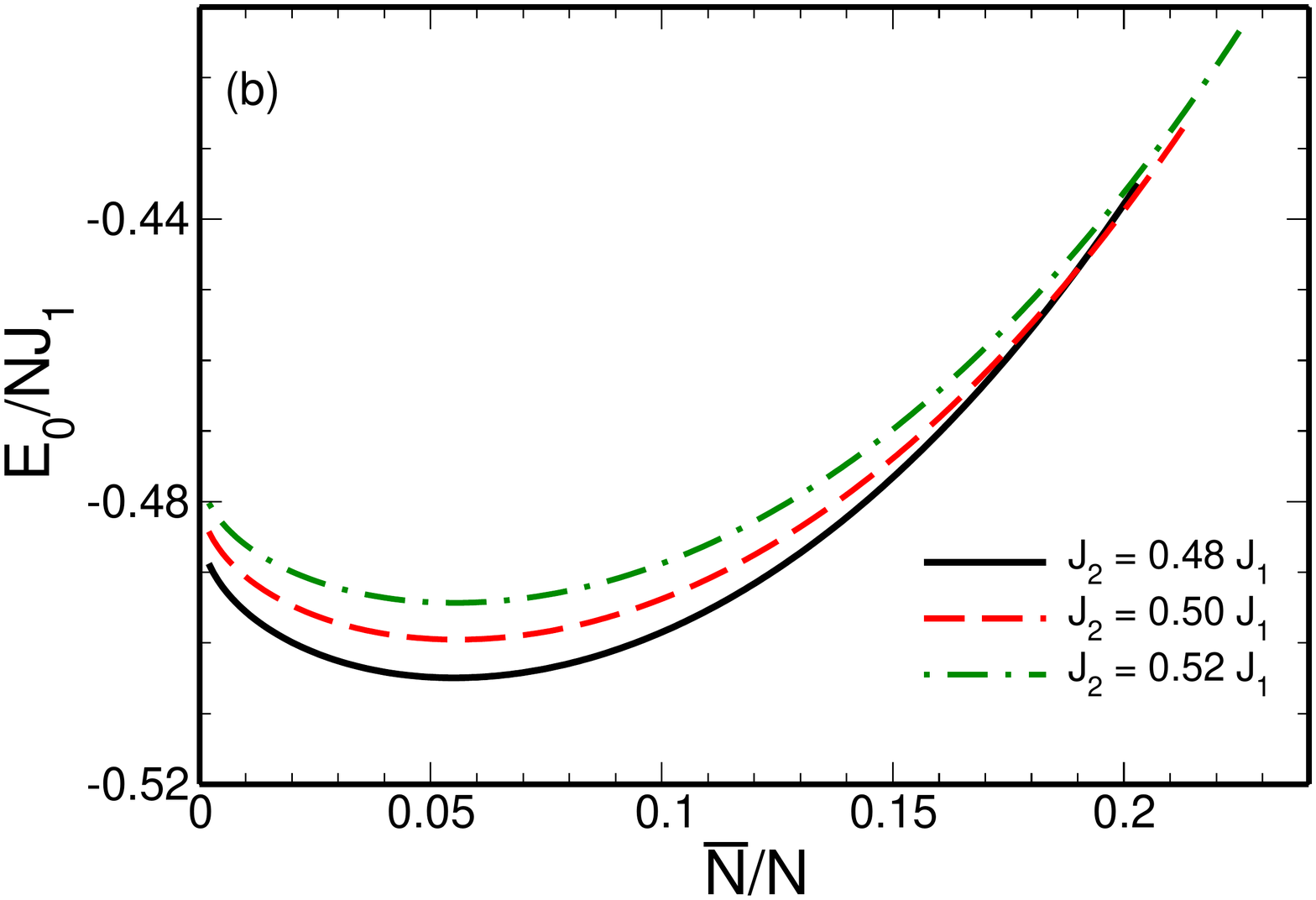}
}
\caption{
  (a) The energies $E_0$ [Eq.~\eqref{egs-bcs}] per site in terms of $J_2/J_1$ of the
  many-triplon state \eqref{mf-wf} with the triplon number 
  $\bar{N} = \bar{N}_{GS}$ (thick solid black line),
  $\bar{N} = 0.10\, N$ (dashed magenta line), and
  $\bar{N} = 0.20\, N$ (dashed green line).
  The corresponding harmonic results for 
  the columnar VBS ground state $E^{\rm HM}_0$ [Eq.~\eqref{egs-harmonic}] (thin solid red line) and for
  the plaquette VBS ground state  (Ref.~\cite{doretto14}) (dotted-dashed blue
  line) are also included.
  (b) The energies $E_0$ [Eq.~\eqref{egs-bcs}] per site in terms of $\bar{N}/N$ of the
  many-triplon state \eqref{mf-wf} for 
  $J_2 = 0.48\,  J_1$ (solid black line),
  $J_2 = 0.50\,  J_1$ (dashed red line), and
  $J_2 = 0.52\, J_1$ (dotted-dashed green line). }
\label{fig:egs}
\end{figure*}

Finally, considering the Fourier transform, 
\begin{equation}
   t_{i \alpha}^\dagger = \frac{1}{ \sqrt{N'} } \sum_{\bk \in {\rm BZ}} 
                  e^{-i\bk \cdot \bR_i } \: t_{\bk \alpha}^\dagger,
\label{fourier}
\end{equation}
where $\bR_i$ is a vector of the dimerized lattice $\mathcal{D}$, 
$N' = N/2$ is the number of dimers ($N$ is the number of sites of the
{\sl original} square lattice), and the momentum sum runs over the
dimerized first Brillouin zone [Fig.~\ref{fig:model}(c)],
we find that, in momentum space, the four terms $\mathcal{H}_n$ 
of the Hamiltonian \eqref{h-effective} read 
\begin{eqnarray}
 \mathcal{H}_0  &=& -\frac{3}{8}J_1NN_0 - \frac{1}{2}\mu N(N_0 - 1),
\label{h0} \\
&& \nonumber \\
\mathcal{H}_2 &=& \sum_\bk \left[ A_\bk t^\dagger_{\bk\alpha}t_{\bk\alpha}
                  + \frac{1}{2}B_\bk \left( t^\dagger_{\bk\alpha}t^\dagger_{-\bk\alpha}
                  + {\rm H.c.}\right) \right],
\label{h2}  \\
&& \nonumber \\
 \mathcal{H}_3 &=& \frac{1}{2\sqrt{N'}}\epsilon_{\alpha\beta\lambda}\sum_{\bp,\bk}\xi_{\bk-\bp}
                  \; t^\dagger_{\bk-\bp\alpha}t^\dagger_{\bp\beta}t_{\bk\lambda} + {\rm H.c.},
\label{h3} \\
&& \nonumber \\
 \mathcal{H}_4 &=& \frac{1}{2N'}\epsilon_{\alpha\beta\lambda}\epsilon_{\alpha\mu\nu}
                  \sum_{\bq,\bp,\bk} \gamma_\bk \;
                   t^\dagger_{\bp+\bk\beta}t^\dagger_{\bq-\bk\mu}t_{\bq\nu}t_{\bp\lambda},
\label{h4}
\end{eqnarray}
with the coefficients $A_\bk$, $B_\bk$, $\xi_\bk$, and $\gamma_\bk$ given by 
\begin{eqnarray}
  A_\bk &=& \frac{1}{4}J_1 - \mu + B_\bk, 
\label{ak} \\
&& \nonumber \\
 B_\bk &=& -\frac{1}{2}N_0\left[ J_1\cos(2k_x) - 2 (J_1 - J_2)\cos(k_y)  \right.
\nonumber \\
&& \left. \right. \nonumber \\
         &+& \left.  J_2\cos(2k_x + k_y) + J_2\cos(2k_x - k_y) \right],       
\label{bk} \\ 
&& \nonumber \\
 \xi_\bk    &=& -\sqrt{N_0}\left[J_1\sin(2k_x) + J_2\sin(2k_x + k_y)  \right.
\nonumber \\
&& \left. \right. \nonumber \\
         &+& \left.  J_2\sin(2k_x - k_y) \right],
\label{xik} \\
&& \nonumber\\
 \gamma_\bk &=& -\frac{1}{2}\left[  J_1\cos(2k_x) + 2(J_1 + J_2)\cos k_y \right.
\nonumber \\
&& \left. \right. \nonumber \\
                    &+& \left. J_2\cos(2k_x + k_y)  +  J_2\cos(2k_x - k_y) \right].
\label{gammak}
\end{eqnarray}

We should mention that the results presented in this section and in
Sec.~\ref{sec:harm} below were previously quoted in Appendix D from
Ref.~\cite{doretto14}. Here, however, we derive them in details
following, e.g., the lines of Ref.~\cite{leite19}.

\begin{figure*}[t]
\centerline{\includegraphics[width=8.5cm]{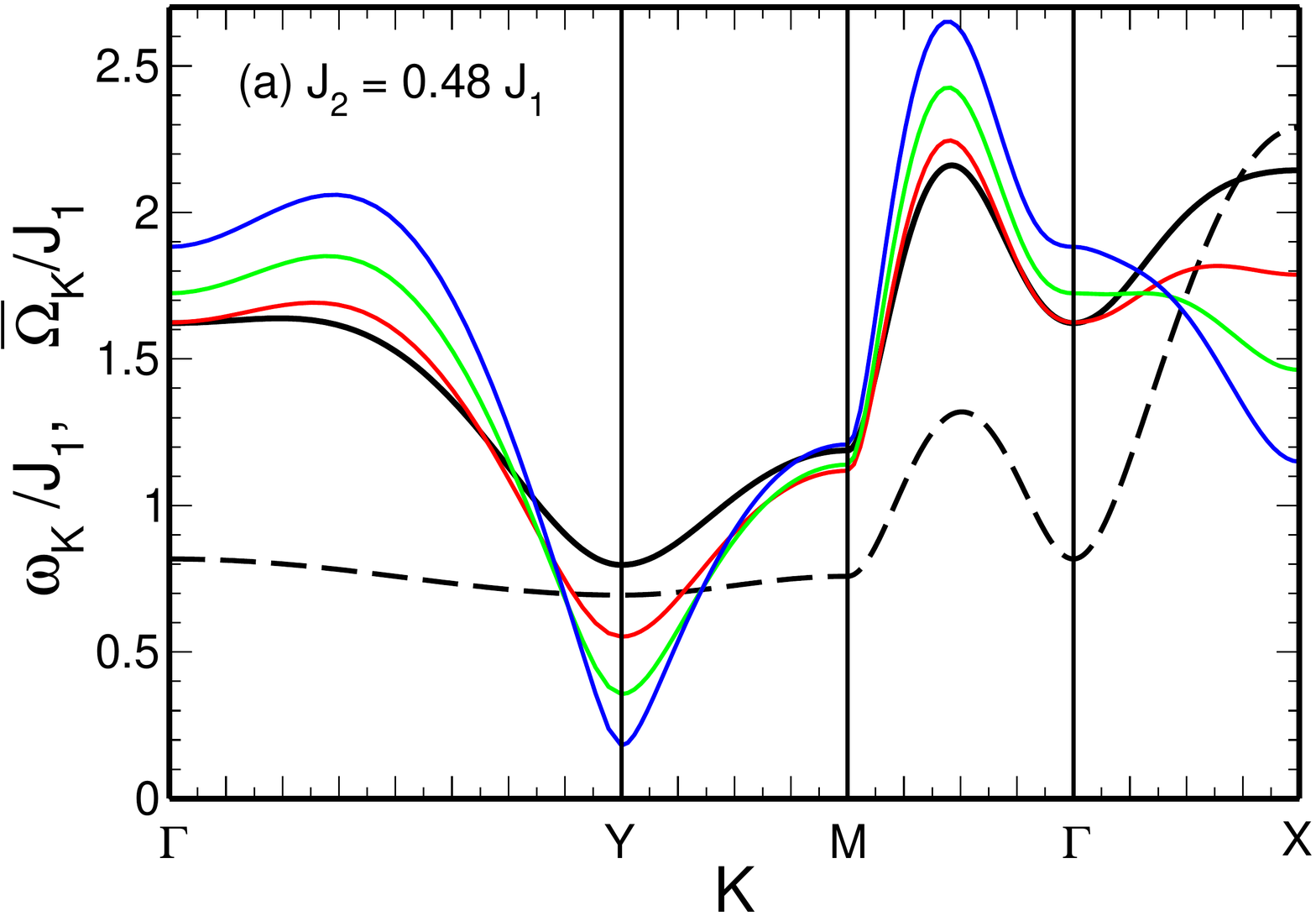}
                   \hskip0.7cm
                  \includegraphics[width=8.5cm]{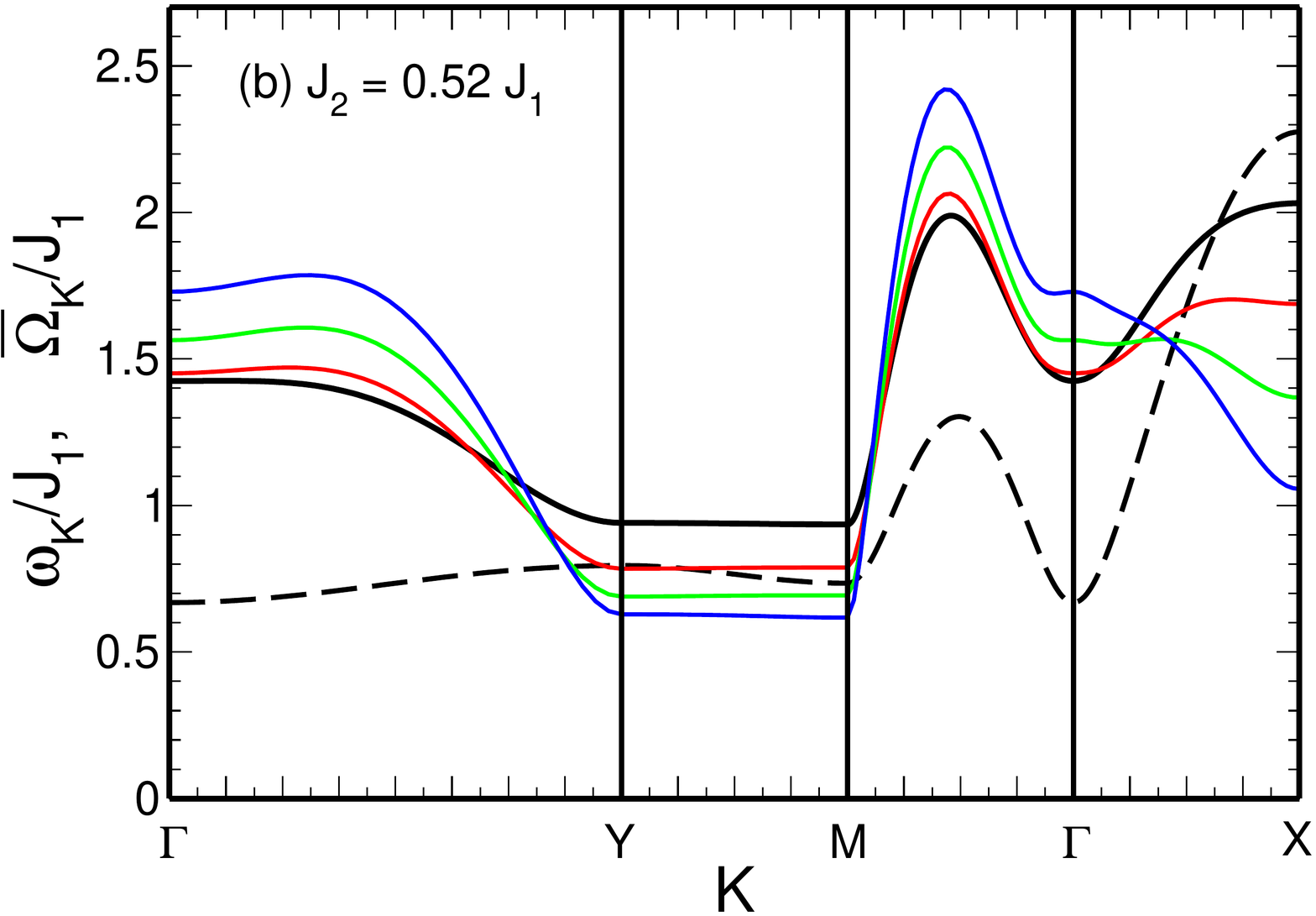}
}
\caption{Triplon dispersion relations $\omega_\bk$
  [Eq.~\eqref{omega-harmonic}] of the columnar VBS phase at the harmonic
  approximation (dashed black  lines) and the dispersion relations
  $\bar{\Omega}_\bk$ [Eq.~\eqref{omega-bcs}] of the elementary
  excitations above the many-triplon state \eqref{mf-wf} 
  within a mean-field approximation (solid lines) along paths in the dimerized
  Brillouin zone [Fig.~\ref{fig:model}(c)] for
  (a) $J_2 = 0.48\, J_1$ and
  (b) $J_2 = 0.52\, J_1$.
  Results for four different values of $\bar{N}$ are shown:
  $\bar{N} = \bar{N}_{GS}$ (solid black line),
  $\bar{N} = 0.10\, N$ (solid red line),
  $\bar{N} = 0.15\, N$ (solid green line), and
  $\bar{N} = 0.20\, N$ (solid blue line).
}
\label{fig:disp}
\end{figure*}

\subsection{Harmonic approximation}
\label{sec:harm}

Let us consider the effective boson model \eqref{h-effective} in the
lowest-order (harmonic) approximation. Here, we only keep the terms of the
Hamiltonian \eqref{h-effective} up to the quadratic order in the
boson operators $t_{\bk\alpha}$,
\begin{equation}
  \mathcal{H}^{\rm HM}_2 \approx \mathcal{H}_0 + \mathcal{H}_2. 
\label{ham-harmonic}
\end{equation}
Since the Hamiltonian $\mathcal{H}^{\rm HM}_2$ is quadratic in the
triplet operators $t_{\bk\alpha}$, it can be diagonalized by the 
Bogoliubov transformation  
\begin{eqnarray}
  t_{\bk\alpha} &=& u_\bk b_{\bk\alpha} - v_\bk b^\dagger_{-\bk\alpha},
\nonumber \\
  t^\dagger_{-\bk\alpha} &=& u_\bk b^\dagger_{-\bk\alpha} - v_\bk b_{\bk\alpha}.
\label{bogo-transf}
\end{eqnarray}
We find that
\begin{equation}
\mathcal{H}^{\rm HM}_ 2 = E^{\rm HM}_0
            + \sum_{\bk\,\alpha} \omega_\bk b^\dagger_{\bk\alpha}b_{\bk\alpha},
\label{diag-ham}
\end{equation}
where 
\begin{equation}
 E^{\rm HM}_0 = -\frac{3}{8}J_1NN_0 - \frac{1}{2}\mu N(N_0 - 1)
     + \frac{3}{2}\sum_\bk \left(\omega_\bk - A_\bk \right)
\label{egs-harmonic}
\end{equation}
is the ground-state energy,
\begin{equation}
 \omega_\bk = \sqrt{A^2_\bk - B^2_\bk}
\label{omega-harmonic}
\end{equation}
is the energy of the {\sl triplons} (the elementary excitations above
the VBS ground state $|{\rm VBS} \rangle$), and the coefficients
$u_\bk$ and $v_\bk$ of the Bogoliubov transformation
\eqref{bogo-transf} read 
\begin{equation}
 u^2_\bk , v^2_\bk = \frac{1}{2}\left( \frac{A_\bk}{\omega_\bk} \pm 1 \right),
 \;\;\;\; {\rm and} \;\;\;\;
 u_\bk v_\bk = \frac{B_\bk}{2\omega_\bk}.
\label{bogo-coef}
\end{equation}
The constants $\mu$ and $N_0$ are calculated from the 
saddle-point conditions 
$\partial E^{\rm HM}_0/\partial N_0 = 0$ and 
$\partial E^{\rm HM}_0/\partial \mu = 0$, and therefore, we find a set of
self-consistent equations,
\begin{eqnarray}
   \mu &=& -\frac{3J_1}{4}   + \frac{3}{2N_0}\frac{1}{N'}\sum_\bk B_\bk \left(  
     \frac{A_\bk - B_\bk}{\omega_\bk}  - 1\right),
\nonumber \\
&& \label{self-mu-nzero} \\
   N_0 &=&  1 - \frac{1}{N'}\sum_{\bk\,\alpha} \langle t^\dagger_{\bk\alpha}t_{\bk\alpha} \rangle
           =   1 + \frac{3}{2N'}\sum_\bk \left(1 - \frac{A_\bk}{\omega_\bk}  \right),  
\nonumber
\end{eqnarray}
that are solved for a fixed value of the ratio $J_2/J_1$.  
Note that, once the constants $\mu$ and $N_0$ are calculated, the
ground-state energy \eqref{egs-harmonic} and 
the triplon dispersion relation \eqref{omega-harmonic}
are completely determined.

The numerical solutions of the set of self-consistent equations
\eqref{self-mu-nzero} are shown in Fig.~\ref{fig:munzero}, where 
the parameters $\mu$ and $N_0$ are plotted as a function of $J_2/J_1$. 
One sees that $N_0$ and $\mu$ monotonically increases with
$J_2/J_1$ up to $J_2 = 0.55\, J_1$.
Moreover, one notices that, within the harmonic approximation, the
columnar VBS phase is stable for $0.30\, J_1 \le  J_2 \le  0.63\, J_1$, 
i.e, a parameter region larger than the one 
($0.4\, J_1 \lesssim J_2 \lesssim 0.6\, J_1$)
expected for the disordered phase of the model \eqref{ham-j1j2}
(see Sec.~\ref{sec:model}).  
Such a feature of the harmonic approximation was found in our previous
studies \cite{doretto12,doretto14,leite19}.

Figure~\ref{fig:egs}(a) shows the ground-state energy
\eqref{egs-harmonic} in terms of the ratio $J_2/J_1$. 
Similar to the parameters $\mu$ and $N_0$, the ground-state energy $E^{\rm HM}_0$
monotonically increases with $J_2/J_1$, but up to $J_2 = 0.57\, J_1$.  
For comparison, the ground-state energy of the plaquette VBS phase
determined within an harmonic approximation (Ref.~\cite{doretto14})
is also included. One sees that, within the corresponding harmonic
approximations, the (tetramerized) plaquette VBS ground state has
lower energy than the (dimerized) columnar VBS one. 
Finally, we should mention that $E^{\rm HM}_0$ of the columnar VBS
ground state was also previously reported in Ref.~\cite{doretto14}. 
However, we found a mistake in our previous numerical code, and
therefore, the results shown in Fig.~\ref{fig:egs}(a) are indeed the
correct ones.

\begin{figure}[t]
\centerline{\includegraphics[width=8.5cm]{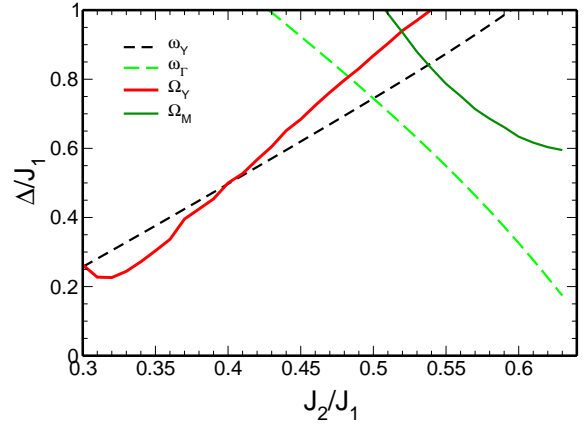} }
\caption{ 
  Harmonic excitation gap (dashed lines) of the columnar VBS phase and the
  excitation gap (solid lines) above the many-triplon state \eqref{mf-wf} with 
  $\bar{N} = \bar{N}_{GS}$ as a function of $J_2/J_1$.  
  $\omega_{\mathbf{Y}}$ (black dashed line) 
  and $\omega_\Gamma$ (green dashed line) are respectively the
  energies $\omega_\bk$ [Eq.~\eqref{omega-harmonic}] of the triplon
  at the $\mathbf{Y} = (0,\pi)$ and the $\Gamma$ points of the
  dimerized Brillouin zone;
  $\bar{\Omega}_{\mathbf{Y}}$ (solid red line) 
  and $\bar{\Omega}_{\mathbf{M}}$ (solid dark green line) are respectively the
  energies $\bar{\Omega}_\bk$ [Eq.~\eqref{omega-bcs}] of the elementary
  excitations above the many-triplon state \eqref{mf-wf} with 
  $\bar{N} = \bar{N}_{GS}$
  at the $\mathbf{Y} = (0,\pi)$ and the $\mathbf{M} = (\pi/2,\pi)$ points
  of the dimerized Brillouin zone.
}
\label{fig:gap}
\end{figure}

The triplon excitation spectrum $\omega_\bk$ [Eq.~\eqref{omega-harmonic}] 
of the columnar VBS phase for $J_2 = 0.48\, J_1$ is shown in
Fig.~\ref{fig:disp}(a).  
As expected for a disordered phase, the triplon excitation spectrum is
gapped. 
Moreover, one notices that the triplon gap (the minimum of the
dispersion relation $\omega_\bk$) is located at the 
$\mathbf{Y} = (0,\pi)$  point of the first Brillouin zone
[Fig.~\ref{fig:model}(c)]. 
Indeed, we find that these two features hold for the 
parameter region $0.30\, J_1 \le J_2 \le 0.50\, J_1$. 
Similarly, for the parameter region  $0.50\, J_1 \le J_2 \le 0.63\,  J_1$, 
we also find a finite triplon excitation gap, but here it is
associated with the $\Gamma$ point (the centre of the first Brillouin
zone), as exemplified in Fig.~\ref{fig:disp}(b) for $J_2 = 0.52\, J_1$.
The complete behaviour of the triplon gap as a function of $J_2/J_1$ 
is shown in Fig.~\ref{fig:gap}.
We should note that, for the first and second parameter regions
above mentioned, 
the momenta associated with the triplon gap are respectively
equal to the ordering wave vectors $\bQ$ of the N\'eel
and collinear magnetic long-range ordered phases that set in for small
and large $J_2$ (see Sec.~\ref{sec:model}). 
As discussed, e.g., in Refs.~\cite{doretto12,doretto14}, the vanishing
of the triplon gap defines a quantum phase transition to a magnetic
ordered phase.
Here, the triplon gap determined within the harmonic approximation
decreases as we approach the critical couplings $J_{2c} = 0.30\, J_1$ and
$J_{2c} = 0.63\, J_1$, but it does not vanish.

As mentioned in Sec.~\ref{sec:intro}, the effects of the cubic [Eq.~\eqref{h3}] 
and the quartic [Eq.~\eqref{h4}] triplet-triplet interactions can be
perturbatively taken into account and corrections to the harmonic
results obtained.  In particular, such effects could decrease the
triplon excitation gap and, eventually, it could vanish at different critical
couplings $J_{2c}$. In this case, the closing of the
triplon gap determines the region of stability of the VBS phase which,
in general, is smaller than the one determined within the harmonic
approximation (see, e.g., Fig.~10 of  Ref.~\cite{doretto12} 
and Fig.~7 of Ref.~\cite{doretto14}). 
Since a carefully determination of the critical couplings $J_{2c}$ 
is not the scope of this paper, we will not employ the
perturbative treatment described above. In the following, we discuss
the many-triplon states based on the harmonic results.

\section{Effective boson model II}
\label{sec:model-b}

Once the triplon operators $b_{\bk\alpha}$ are defined in terms of the
triplet operators $t_{\bk\alpha}$ [Eq.~\eqref{bogo-transf}] and the
triplon spectrum \eqref{omega-harmonic} and the triplon vacuum 
$|{\rm VBS}\rangle$ are determined within the
harmonic approximation, we now consider a 
system with a fixed number $\bar{N}$ of triplons.

We start expressing the effective boson model \eqref{h-effective}
in terms of the boson operators $b_{\bk\alpha}$, i.e., we
derive an effective boson model for the triplons $b$.
With the aid of the Bogoliubov transformation \eqref{bogo-transf}, one
shows that the cubic term \eqref{h3} can be written in
terms of the $b$ operators as \cite{doretto12} 
\begin{eqnarray}
\mathcal{H}_3 &=& \frac{1}{2\sqrt{N'}}\sum_{\bk,\bp}
                   \sideset{}{'}\sum_{\alpha,\beta,\gamma}
                   \Xi_1(\bk,\bp)
               (b^\dagger_{\bk-\bp \alpha}b^\dagger_{\bp \beta}b_{\bk \gamma}
                       + {\rm H.c.})
\nonumber \\
             &+& \frac{1}{2\sqrt{N'}}\sum_{\bk,\bp}
                 \Xi_2(\bk,\bp)
                   (b^\dagger_{\bk-\bp x}b^\dagger_{\bp y}b^\dagger_{-\bk z}
                 + {\rm H.c.}).
\label{h3-b}
\end{eqnarray}
Here, the sum over $\alpha,\beta,\gamma$ has only three components,
$(\alpha,\beta,\gamma) = (x,y,z)$, $(z,x,y)$, and $(y,z,x)$,
the renormalized vertex $\Xi_1(\bk,\bp)$ reads
\begin{eqnarray}
\Xi_1(\bk,\bp) &=&
   \left(\xi_{\bk-\bp} - \xi_\bp\right)\left(u_{\bk-\bp}u_\bp u_\bk
                      +  v_{\bk-\bp}v_\bp v_\bk \right)
\nonumber \\
        &+& \left(\xi_\bk + \xi_\bp\right)\left(v_{\bk-\bp}u_\bp v_\bk
                      +  u_{\bk-\bp}v_\bp u_\bk \right)
\nonumber \\
        &-&  \left(\xi_{\bk-\bp} + \xi_\bk\right)\left(v_{\bk-\bp}u_\bp u_\bk
                      +  u_{\bk-\bp}v_\bp v_\bk
                      \right),\;\;\;\;
\label{gammas}
\end{eqnarray}
with $\xi_\bk$ being the bare cubic vertex \eqref{xik} and
$u_\bk$ and $v_\bk$ being the Bogoliubov coefficients \eqref{bogo-coef}.
The vertex $\Xi_2(\bk,\bp) = - \Xi_1(\bk,\bp)$ with the
replacements $u_\bk \leftrightarrow v_\bk$.

Following the same procedure for the quartic term \eqref{h4}, one
shows, after normal-ordering, that   
\begin{equation}
 \mathcal{H}_4 = E_{40} + \mathcal{H}_{24} + \mathcal{H}_{44}, 
\label{h4-b}
\end{equation}
where
\begin{equation}
 E_{40}  = \frac{3}{N'} \sum_{\bk,\bp} \gamma_{\bk - \bp} 
                       \left(  u_\bk v_\bk u_\bp v_\bp -  v^2_\bk v^2_\bp \right),
\label{h40-b} 
\end{equation}
\begin{equation}
\mathcal{H}_{24} = \sum_\bk \left[ A^{(4)}_\bk b^\dagger_{\bk\alpha}b_{\bk\alpha}
                  + \frac{1}{2}B^{(4)}_\bk \left( b^\dagger_{\bk\alpha}b^\dagger_{-\bk\alpha}
                  + {\rm H.c.}\right) \right],
\label{h24-b}
\end{equation}
and
\begin{eqnarray}
\mathcal{H}_{44} &=& \frac{1}{2N'}\epsilon_{\alpha\beta\lambda}\epsilon_{\alpha\mu\nu}
                  \sum_{\bq,\bp,\bk} \left[   \right. 
\nonumber \\
&& \nonumber \\
                  && \left. \Gamma_1(\bp,\bq,\bk) \;
                   b^\dagger_{\bp+\bk\beta} b^\dagger_{\bq-\bk\mu} b^\dagger_{-\bq\nu} b^\dagger_{-\bp\lambda}
                   + {\rm H.c.} \right.
\nonumber \\
&& \nonumber \\
        && + \left. \Gamma_2(\bp,\bq,\bk) \;
        b^\dagger_{\bp+\bk\beta} b^\dagger_{\bq-\bk\mu} b^\dagger_{-\bq\nu} b_{\bp\lambda}
        + {\rm H.c.} \right.
\nonumber \\
&& \nonumber \\
        && + \left. \Gamma_3(\bp,\bq,\bk) \;
        b^\dagger_{\bp+\bk\beta} b^\dagger_{\bq-\bk\mu} b_{\bq\nu} b_{\bp\lambda} \right.
\nonumber \\
&& \nonumber \\
        && + \left. \Gamma_4(\bp,\bq,\bk) \;
        b^\dagger_{\bp+\bk\beta} b^\dagger_{-\bp\lambda} b_{-\bq+\bk\mu} b_{\bq\nu} \right].
\label{h44-b}
\end{eqnarray}
Here, the coefficients $A^{(4)}_\bk$ and $B^{(4)}_\bk$ are given by
\begin{eqnarray}
 A^{(4)}_\bk   &=& \frac{2}{N'} \sum_\bp \gamma_{\bk - \bp} 
                  \left[  2u_\bk v_\bk  u_\bp v_\bp  -  ( u^2_\bk  +  v^2_\bk ) v^2_\bp  \right],
\nonumber \\
&& \label{ab4k} \\
 B^{(4)}_\bk   &=& \frac{2}{N'} \sum_\bp \gamma_{\bk - \bp} 
                  \left[  2u_\bk v_\bk v^2_\bp   - ( u^2_\bk   + v^2_\bk ) u_\bp v_\bp  \right],
\nonumber
\end{eqnarray}
and the functions $\Gamma_i(\bp,\bq,\bk)$ read
\begin{eqnarray}
  \Gamma_1(\bp,\bq,\bk) &=& \gamma_\bk \, u_{\bp+\bk} u_{\bq -\bk} v_\bq v_\bp,
\nonumber \\
&& \nonumber \\
  \Gamma_2(\bp,\bq,\bk) &=& 2\gamma_\bk 
     \left( v_{\bp+\bk} u_{\bq - \bk} v_\bq v_\bp  -  u_{\bp+\bk} u_{\bq - \bk} v_\bq u_\bp  \right),
\nonumber \\
&& \nonumber \\
  \Gamma_3(\bp,\bq,\bk) &=& \gamma_\bk 
                              \left(  u_{\bp+\bk} u_{\bq - \bk} u_\bq u_\bp 
                                     + v_{\bp+\bk} v_{\bq - \bk} v_\bq v_\bp \right.
 \nonumber \\
&& \nonumber \\
             && \left.          - u_{\bp+\bk} v_{\bq - \bk} v_\bq u_\bp
                                     - v_{\bp+\bk} u_{\bq - \bk} u_\bq v_\bp \right),
\nonumber \\
&& \nonumber \\
  \Gamma_4(\bp,\bq,\bk) &=& 2\gamma_\bk \, u_{\bp+\bk} v_{\bq - \bk} u_\bq v_\bp,
\label{gi-functions}
\end{eqnarray}
with $\gamma_\bk$ being the bare quartic vertex \eqref{gammak}
and $u_\bk$ and $v_\bk$, the Bogoliubov coefficients \eqref{bogo-coef}.
We refer the reader to Eq.~\eqref{ab4k02} for alternative expressions
for the constant $E_{40}$ and the coefficients $A^{(4)}_\bk$ and $B^{(4)}_\bk$  
that are useful in the self-consistent problem discussed in the next
section.

Therefore, the effective (interacting) boson model for the triplons $b$ 
(considering the columnar VBS as a reference state) assumes the form
\begin{equation}
 \mathcal{H} = \mathcal{H}^{\rm HM}_2 + \mathcal{H}_3 + \mathcal{H}_4, 
\label{h-effective2}
\end{equation}
where $\mathcal{H}^{\rm HM}_2$ is the (quadratic) harmonic Hamiltonian
\eqref{diag-ham}, and the cubic $\mathcal{H}_3$ and the quartic $\mathcal{H}_4$
terms are respectively given by Eqs.~\eqref{h3-b} and \eqref{h4-b}.

\subsection{Mean-field approximation}
\label{sec:mf}

In this section, we study systems with a fixed number $\bar{N}$ of triplons
described by the Hamiltonian \eqref{h-effective2}. In particular,
we neglected the cubic term $\mathcal{H}_3$ and consider the quartic
term $\mathcal{H}_4$ within a mean-field approximation. 
The idea is to verify whether a ground state formed by a certain
number of triplons (the {\sl many-triplon state}) is stable, 
in addition to determine the corresponding excitation spectrum.

It is easy to show that, within a mean-field approximation, the quartic
term $\mathcal{H}_{44}$ [Eq.~\eqref{h44-b}] assumes the form
\begin{eqnarray}
\mathcal{H}^{MF}_{44} = E_{44} 
   + \sum_{\bk\alpha} && \left[  \Delta_{1,\bk} b^\dagger_{\bk\alpha}b_{\bk\alpha}
      \right.
\nonumber \\           
   && \left.  +  \frac{1}{2} \Delta_{2,\bk} \left( b^\dagger_{\bk\alpha}b^\dagger_{-\bk\alpha}
                   + {\rm H.c.}\right) \right],
\label{h44-b-mf}
\end{eqnarray}
where the constant $E_{44}$ [Eq.~\eqref{e44}] and the coefficients 
$\Delta_{1,\bk}$ [Eq.~\eqref{delta1}] and $\Delta_{2,\bk}$
[Eq.~\eqref{delta2}] are defined in terms of the bare quartic vertex
$\gamma_\bk$ [Eq.~\eqref{gammak}], the Bogoliubov coefficients
$u_\bk$ and $v_\bk$ [Eq.~\eqref{bogo-coef}], and the normal ($h_\bk$) and
anomalous ($\bar{h}_\bk$) expectation values:   
\begin{equation}
h_\bk \equiv \langle b^\dagger_{\bk\alpha} b_{\bk\alpha} \rangle,
\;\;\;\;\;\;\;\;\;\;\;
\bar{h}_\bk \equiv \langle b_{\bk\alpha} b_{-\bk\alpha} \rangle.
\label{mf-par}
\end{equation}
Due to the fact that the quartic term  $\mathcal{H}_{44}$ does not
conserve the number of particles, one should include not only $h_\bk$
but also $\bar{h}_\bk$. Moreover, we consider both normal and
anomalous expectation values $\alpha$ independent, since the triplon
energy \eqref{omega-harmonic} and the quartic vertices
$\Gamma_i(\bp,\bq,\bk)$ [Eq.~\eqref{gi-functions}] do not depend on the
index $\alpha$.

From Eqs.~\eqref{diag-ham}, \eqref{h40-b}, \eqref{h24-b}, and
\eqref{h44-b-mf}, we then find that the mean field Hamiltonian for a
system of $\bar{N}$ triplons is given by
\begin{eqnarray}
\mathcal{H} &=& \mathcal{H}^{MF} - \bar{\mu}\bar{N} 
\nonumber \\
&& \nonumber \\
\mathcal{H} &=& \mathcal{H}_2^{HM} + E_{40}  + \mathcal{H}_{24} +\mathcal{H}_{44}^{MF} 
  - \bar{\mu} \bar{N}  
\nonumber \\    
&& \nonumber \\
\mathcal{H} &=& E^{HM}_0  + E_{40}  +  E_{44}   
\nonumber \\
&& \nonumber \\
    &+& \sum_{\bk\alpha} \left[ \bar{A}_\bk b^\dagger_{\bk\alpha}b_{\bk\alpha}
                  + \frac{1}{2} \bar{B}_\bk \left( b^\dagger_{\bk\alpha}b^\dagger_{-\bk\alpha}
                  + {\rm H.c.}\right) \right].
\label{h-mf}
\end{eqnarray}
Here, the coefficients $\bar{A}_\bk$ and $\bar{B}_\bk$ read
\begin{eqnarray}
 \bar{A}_\bk &=& \omega_\bk + A^{(4)}_\bk + \Delta_{1,\bk} - \bar{\mu},
\nonumber \\
&& \nonumber \\
 \bar{B}_\bk &=& B^{(4)}_\bk + \Delta_{2,\bk},
\label{abbark}
\end{eqnarray}
with $\omega_\bk$ being the harmonic triplon energy \eqref{omega-harmonic}, 
the coefficients $A^{(4)}_\bk$ and $B^{(4)}_\bk$ given by Eq.~\eqref{ab4k},
and the coefficients $\Delta_{1,\bk}$ and $\Delta_{2,\bk}$
respectively given by Eqs.~\eqref{delta1} and \eqref{delta2}.
Moreover,  $\bar{\mu} $ is the chemical potential related to the total
number of triplons $b$, i.e.,
\begin{equation}
 \bar{N} = \sum_{\bk\,\alpha}  b^\dagger_{\bk\alpha}b_{\bk\alpha}. 
\label{constraint2}
\end{equation}

The Hamiltonian \eqref{h-mf} can be diagonalized by a 
Bogoliubov transformation similar to the transformation
\eqref{bogo-transf}:
\begin{eqnarray}
  b_{\bk\alpha} &=& \bar{u}_\bk a_{\bk\alpha} - \bar{v}_\bk a^\dagger_{-\bk\alpha},
\nonumber \\ 
  b^\dagger_{-\bk\alpha} &=& \bar{u}_\bk a^\dagger_{-\bk\alpha} - \bar{v}_\bk a_{\bk\alpha}.
\label{bogo-transf2}
\end{eqnarray}
We then arrive at
\begin{equation}
 \mathcal{H} = E_0 + \sum_{\bk\,\alpha} \bar{\Omega}_\bk a^\dagger_{\bk\alpha} a_{\bk\alpha},
\end{equation}
where
\begin{equation}
 E_0 = E^{HM}_0  + E_{40}  +  E_{44}   
     + \frac{3}{2}\sum_\bk \left( \bar{\Omega}_\bk - \bar{A}_\bk \right)
\label{egs-bcs}
\end{equation}
is the energy of the many-triplon state $|\Psi_0\rangle$ [see
Eq.~\eqref{mf-wf} below],
\begin{equation}
   \bar{\Omega}_\bk = \sqrt{ \bar{A}^2_\bk - \bar{B}^2_\bk }
\label{omega-bcs}
\end{equation}
is the energy of the elementary excitations above the many-triplon state
$|\Psi_0\rangle$, and the coefficients $\bar{u}_\bk$ and $\bar{v}_\bk$
of the Bogoliubov transformation \eqref{bogo-transf2} are given by
\begin{equation}
  \bar{u}^2_\bk , \bar{v}^2_\bk  = \frac{1}{2} \left( \frac{\bar{A}_\bk }{\bar{\Omega}_\bk } \pm 1 \right)
 \;\;\;\; {\rm and} \;\;\;\;
 \bar{u}_\bk \bar{v}_\bk = \frac{ \bar{B}_\bk }{ 2\bar{\Omega}_\bk }.
\label{bogo-coef-bcs}
\end{equation}

\begin{figure}[t]
\centerline{\includegraphics[width=8.5cm]{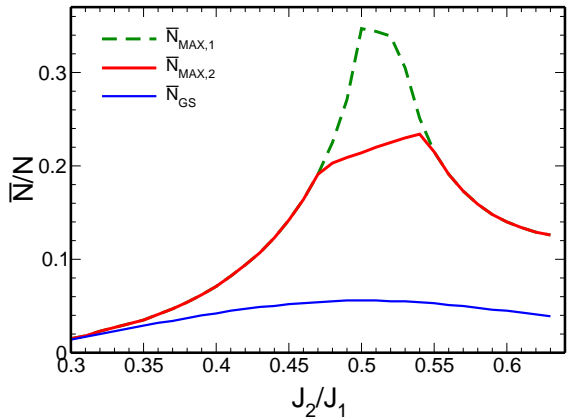}}
\caption{The parameters 
  $\bar{N}_{MAX,1}$ (dashed green line), 
  $\bar{N}_{MAX,2}$ (thick solid red line), and
  $\bar{N}_{GS}$ (thin solid blue line) as a function of $J_2/J_1$
  determined from the numerical solutions of the self-consistent
  problem \eqref{defb1}-\eqref{nbar02}.}
\label{fig:nbarmax}
\end{figure}

From Eqs.~\eqref{bogo-transf2} and \eqref{bogo-coef-bcs}, one shows
that the normal $h_\bk$ and anomalous $\bar{h}_\bk$ expectation values
\eqref{mf-par} assume the form
\begin{eqnarray}
 h_\bp &=& \langle b^\dagger_{\bp\alpha} b_{\bp\alpha} \rangle
                     = \bar{v}^2_\bp
                     = \frac{1}{2} \left( -1 + \frac{ \bar{A}_\bk}{\Omega_\bk } \right),
\nonumber \\
\bar{h}_\bp &=& \langle b_{\bp\alpha} b_{-\bp\alpha} \rangle 
                       = - \bar{v}_\bp \bar{u}_\bp 
                       =  - \frac{ \bar{B}_\bp }{ 2\Omega_\bp }.
\label{mf-par2}
\end{eqnarray}
Moreover, considering the condition \eqref{constraint2} on average,
Eq.~\eqref{mf-par2} yields  
\begin{equation}
  \frac{\bar{N}}{N} = \frac{3}{4N'} \sum_\bk  
      \left( -1 + \frac{\bar{A}_\bp}{\Omega_\bp} \right), 
\label{nbar}
\end{equation}
where $N$ is the number of sites of the {\sl original} square lattice.
Important, only systems with $\bar{N} \le N/2$ should be
considered.

We determine the normal and anomalous expectation values
\eqref{mf-par2} and the chemical potential $\bar{\mu}$ related to the
condition \eqref{nbar} for fixed values of the triplon number
$\bar{N}$ and the ratio $J_2/J_1$ of the exchange couplings
by numerically solving the self-consistent problem defined by
Eqs.~\eqref{defb1}-\eqref{nbar02}. 
We refer the reader to Appendix~\ref{ap:details-mf} 
for the details of the self-consistent procedure. 
Important, for a given value of the ratio $J_2/J_1$, we consider
the values of the parameter $N_0$ and the Lagrange multiplier $\mu$ 
determined within the harmonic approximation  
for the columnar VBS state [Fig.~\ref{fig:munzero}], since it is the
reference state that defines the triplons $b$.

Before discussing the numerical results, a few remarks here about the
nature of the many-triplon state $|\Psi_0\rangle$ are in order: It is
possible to show that the expectation values \eqref{mf-par2} are
consistent with the state 
\begin{equation}
 | \Psi_0 \rangle = {\rm C} \prod_\bk \exp\left( 
     -\phi_\bk b^\dagger_{-\bk\alpha} b^\dagger_{\bk\alpha}  \right)| \rm{VBS} \rangle,
\label{mf-wf}
\end{equation}
where $\phi_\bk = \bar{v}_\bk/\bar{u}_\bk$, with $\bar{u}_\bk$ and
$\bar{v}_\bk$ being the Bogoliubov coefficients \eqref{bogo-coef-bcs}, 
the normalization constant ${\rm C}^{-2} = \prod_\bk \bar{u}^2_\bk$, and 
$|\rm{VBS} \rangle$ is the vacuum for the triplons $b$.
Therefore, the many-triplon state (within our mean-field
approximation) is a BCS-like
state that correlates pairs of triplons $b$ with momenta $\bk$ and
$-\bk$ and the same index $\alpha = x, y, z$. 
Important, $| \Psi_0 \rangle$ does not describe a triplon-pair
condensate, since here there is no  $U(1)$ symmetry to be broken.  
Indeed, both the Hamiltonian \eqref{h-effective2} and the ground
state \eqref{mf-wf} only preserve a Z$_2$ symmetry: 
$b_\bk \rightarrow -b_\bk$.

For a fixed value of the ratio $J_2/J_1$, we find numerical solutions for the
self-consistent problem \eqref{defb1}-\eqref{nbar02} only for
$\bar{N} \le \bar{N}_{MAX,1}$, where the values of the parameter
$\bar{N}_{MAX,1}$ as a function of $J_2/J_1$ are shown in Fig.~\ref{fig:nbarmax}.  
Important, for $0.48 J_1 \le J_2 \le 0.54 J_1$, we find solutions for
the self-consistent problem with 
\[
   N_{\rm Triplet} = \frac{1}{N'}\sum_{\bk\,\alpha} 
                     \langle t^\dagger_{\bk\alpha}t_{\bk\alpha} \rangle > 1.
\]
Taking into account the additional condition $N_{\rm Triplet}<1$, we
define the parameter $\bar{N}_{MAX,2}$ (see Fig~\ref{fig:nbarmax}), 
and therefore, we only consider solutions of the
self-consistent problem with $\bar{N} \le \bar{N}_{MAX,2}$.

\begin{figure*}[t]
\centerline{\includegraphics[width=8.5cm]{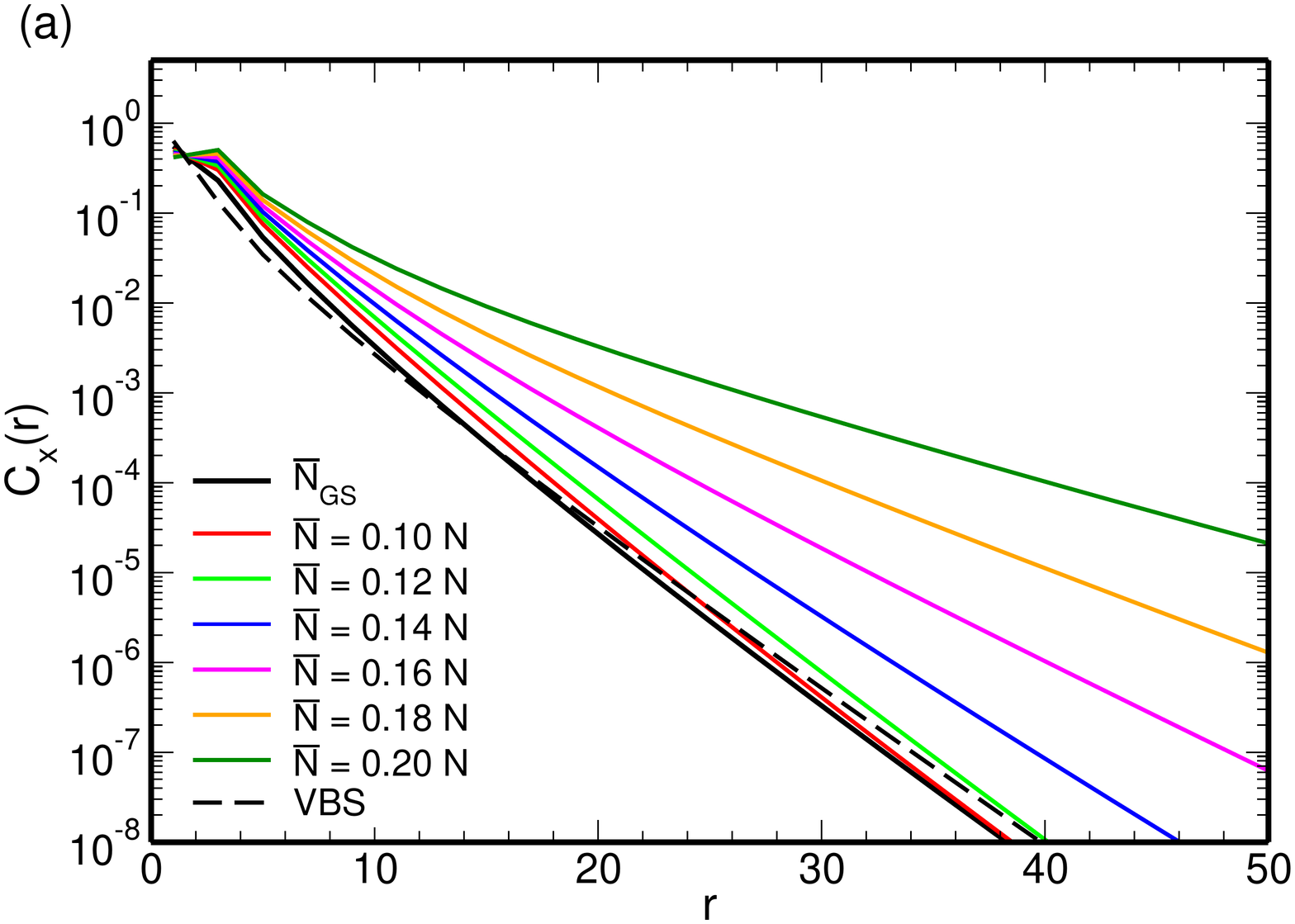}
                   \hskip0.5cm
                  \includegraphics[width=8.5cm]{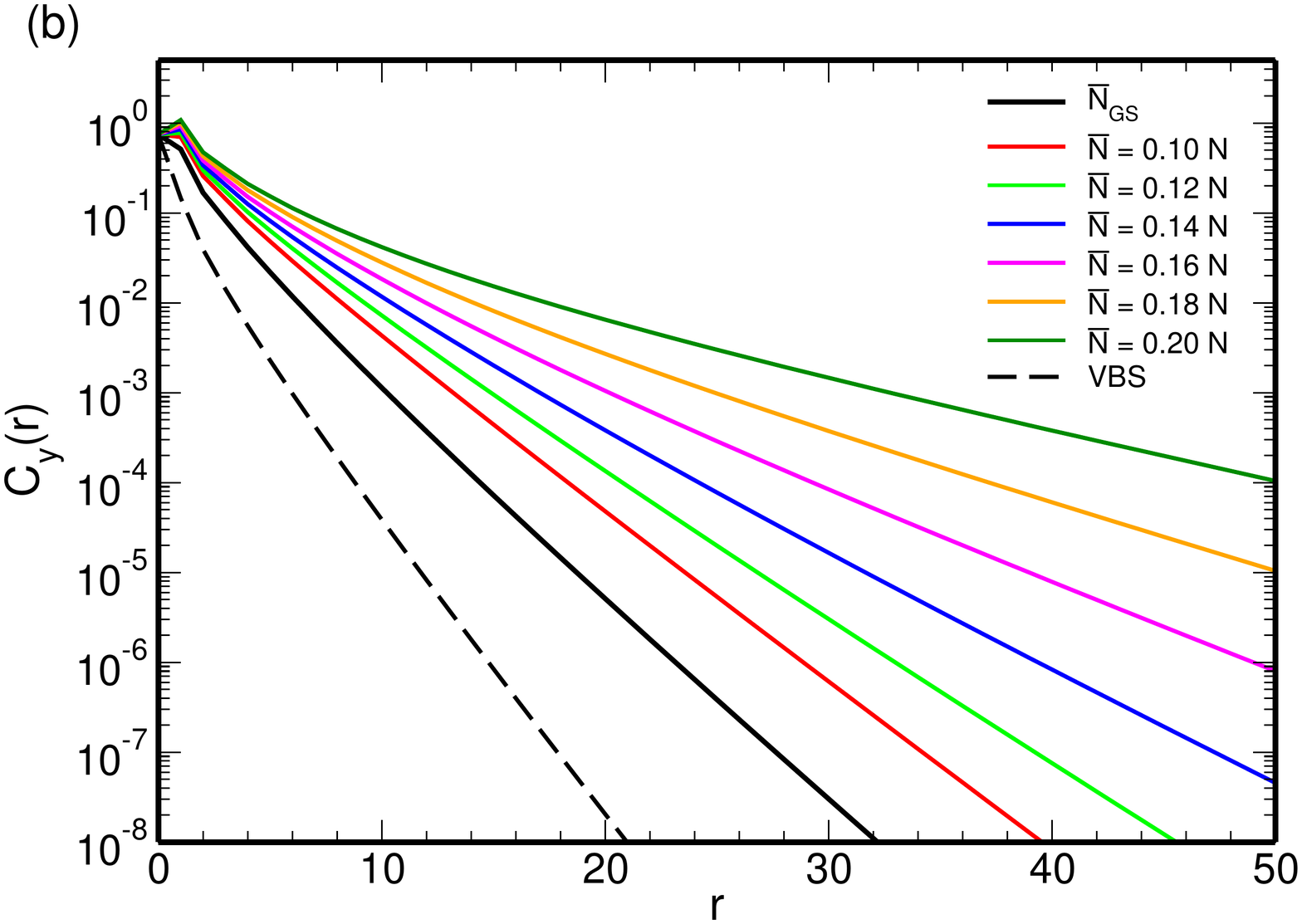}
}
\caption{Spin-spin correlation functions (odd distances $r$)  
  (a) $C_x(r)$ [Eq.~\eqref{cor-spin-x}] and
  (b) $C_y(r)$ [Eq.~\eqref{cor-spin-y}] of the many-triplon state \eqref{mf-wf}
  for $J_2 = 0.48\, J_1$. Mean-field results for different values 
  of the triplon number $\bar{N}$ (solid lines) are shown:
  $\bar{N} = \bar{N}_{GS}$ (black),
  $\bar{N} = 0.10\, N$ (red),
  $\bar{N} = 0.12\, N$ (green),
  $\bar{N} = 0.14\, N$ (blue),
  $\bar{N} = 0.16\, N$ (magenta),
  $\bar{N} = 0.18\, N$ (orange), and
  $\bar{N} = 0.20\, N$ (dark green).
  The corresponding harmonic results for the columnar VBS ground-state $|{\rm VBS}\rangle$ 
  are also included (dashed black line). }
\label{fig:cor-spin01}
\end{figure*}

For a given value of the ratio $J_2/J_1$ of the exchange couplings, the
energy $E_0$ [Eq.~\eqref{egs-bcs}] of the many-triplon state \eqref{mf-wf} 
has a non-monotonic behaviour as $\bar{N}$ increases
[Fig.~\ref{fig:egs}(b)] and, in particular, it reaches a 
minimum value at $\bar{N} =\bar{N}_{GS}$, where the values of $\bar{N}_{GS}$ as a
function of $J_2/J_1$ are displayed in Fig.~\ref{fig:nbarmax}.  
One notices that $\bar{N}_{GS} < 0.056\, N$, where $N$ is the number of
sites of the original square lattice, i.e, the lowest-energy
many-triplon state \eqref{mf-wf} has a small number of
triplons $b$.  
The behaviour of the energy \eqref{egs-bcs} for 
$\bar{N} =\bar{N}_{GS}$ as a function of $J_2/J_1$ is shown in
Fig.~\ref{fig:egs}(a). 
Interesting, one sees that, for a given value of $J_2/J_1$, the
ground-state energy $E_0$ of the many-triplon state with 
$\bar{N}=\bar{N}_{GS}$ is smaller than the ones of the columnar
[Eq.~\eqref{egs-harmonic}]  and plaquette (Ref.~\cite{doretto14}) VBSs
both determined at the corresponding harmonic levels.

Figure~\ref{fig:disp} shows the energy of the elementary
excitations $\bar{\Omega}_\bk$ [Eq.~\eqref{omega-bcs}] above the
many-triplon state \eqref{mf-wf} with $\bar{N} =\bar{N}_{GS}$ 
for $J_2 = 0.48 J_1$ [Fig.~\ref{fig:disp}(a)] and 
$J_2 = 0.52 J_1$ [Fig.~\ref{fig:disp}(b)]. 
Apart from the momenta close to the $X$ point, one sees that, for both
values of the model parameter $J_2$, $\bar{\Omega}_\bk > \omega_\bk$,
where $\omega_\bk$ is the corresponding (harmonic) triplon energy
\eqref{omega-harmonic}.  
In particular, for $J_2=0.48 J_1$, the excitation gap is located at the $Y$ point, 
the same momentum associated with the triplon gap of the corresponding
columnar VBS state [dashed black line, Fig.~\ref{fig:disp}(a)]. 
On the other hand, for $J_2 = 0.52 J_1$, the excitation gap is located
at the $M$ point, different from the corresponding columnar VBS state
whose triplon gap is associated with the $\Gamma$ point
[dashed black line, Fig.~\ref{fig:disp}(b)]. 
Indeed, one finds that the features described above for $J_2 = 0.48 J_1$ 
hold for the parameter region $J_2 \le 0.51 J_1$, while the ones found
for $J_2 = 0.52 J_1$, for the parameter region $J_2 > 0.51 J_1$.
The complete behaviour of the excitation gap above the many-triplon 
state with $\bar{N} =\bar{N}_{GS}$ as a function of $J_2/J_1$ and a
comparison with the (harmonic) excitation gap of the columnar VBS
ground state are shown in Fig.~\ref{fig:gap}.

In addition to the lowest-energy many-triplon state \eqref{mf-wf} with 
$\bar{N} = \bar{N}_{GS}$, we also consider (high-energy) triplon 
states with larger number of triplons, $\bar{N} > \bar{N}_{GS}$. 
In particular, in Fig.~\ref{fig:egs}(a), we show the energy \eqref{egs-bcs}  
of the many-triplon state for $\bar{N} = 0.10\,N$ and $0.20\,N$ 
in terms of $J_2/J_1$. As already mentioned, for a given value of the ratio
$J_2/J_1$, the energy \eqref{egs-bcs}  
increases with the number of triplons $\bar{N}$ when $\bar{N} > \bar{N}_{GS}$.  
Figure~\ref{fig:disp} also displays the spectra of the elementary
excitations $\bar{\Omega}_\bk$ [Eq.~\eqref{omega-bcs}] above the
many-triplon state \eqref{mf-wf}
for $\bar{N} = 0.10\,N$, $0.15\,N$, and $0.20\,N$
and $J_2 = 0.48  J_1$ [Fig.~\ref{fig:disp}(a)]
and $J_2 = 0.52 J_1$ [Fig.~\ref{fig:disp}(b)]. 
Apart from the region around the $X$ point, the excitation spectra
for $\bar{N} > \bar{N}_{GS}$ have the same (qualitatively) features of the
corresponding ones for $\bar{N} = \bar{N}_{GS}$. Moreover, 
we notice that, as $\bar{N}$ increases from $\bar{N}_{GS}$ to $0.20\, N$,
the excitation gap decreases and, in particular, 
it decreases faster for $J_2 = 0.48  J_1$ than for 
$J_2 = 0.52 J_1$.  Indeed, the excitation gap almost vanishes as
$\bar{N}$ approaches $\bar{N}_{MAX,1}$. 
However, recall that, for the region $0.48 J_1 \le J_2 \le 0.54 J_1$,   
we should only consider solutions of the self-consistent problem
\eqref{defb1}-\eqref{nbar02}  with $\bar{N} \le \bar{N}_{MAX,2}$.

\begin{figure*}[t]
\centerline{\includegraphics[width=8.5cm]{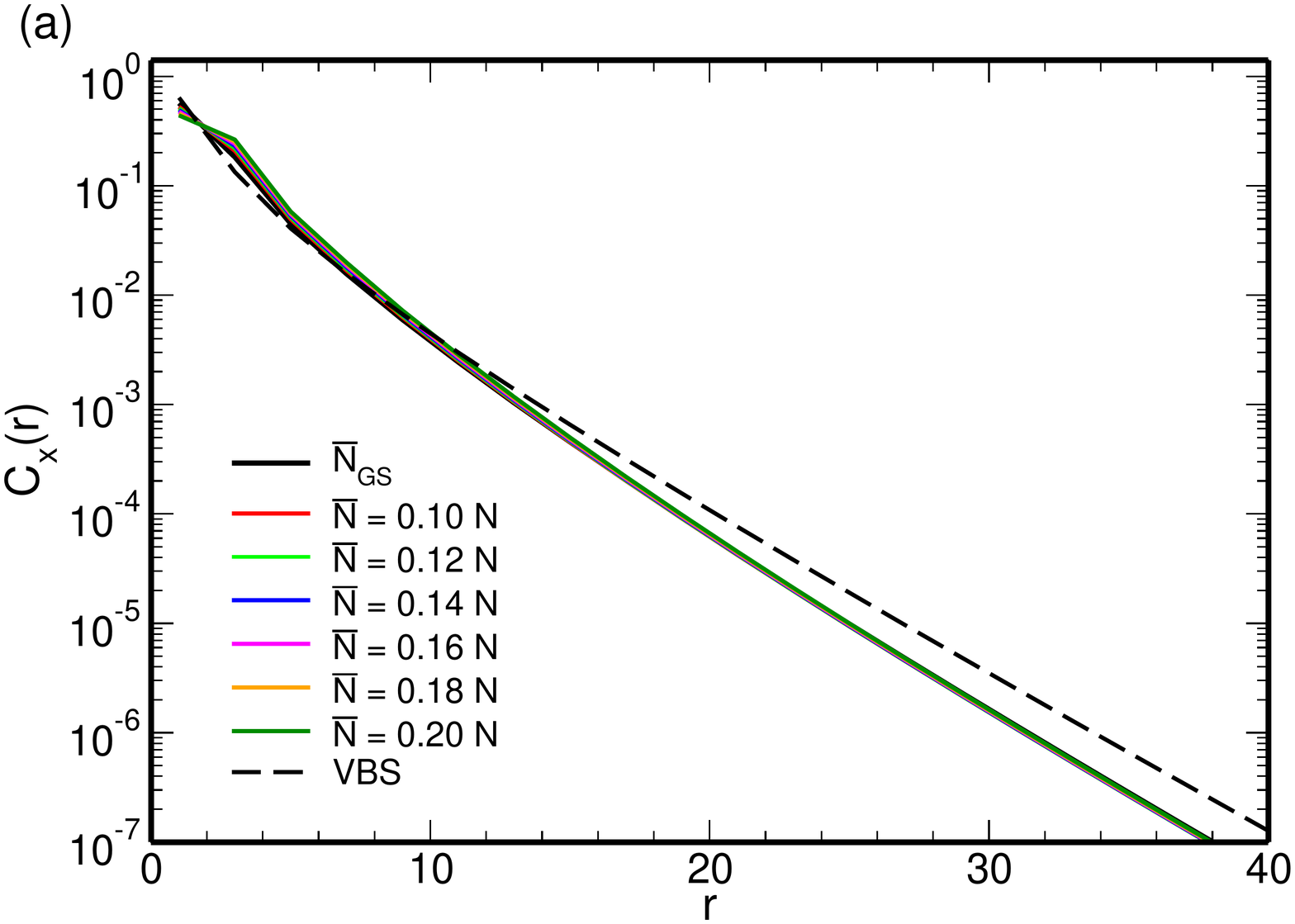}
                   \hskip0.5cm
                  \includegraphics[width=8.5cm]{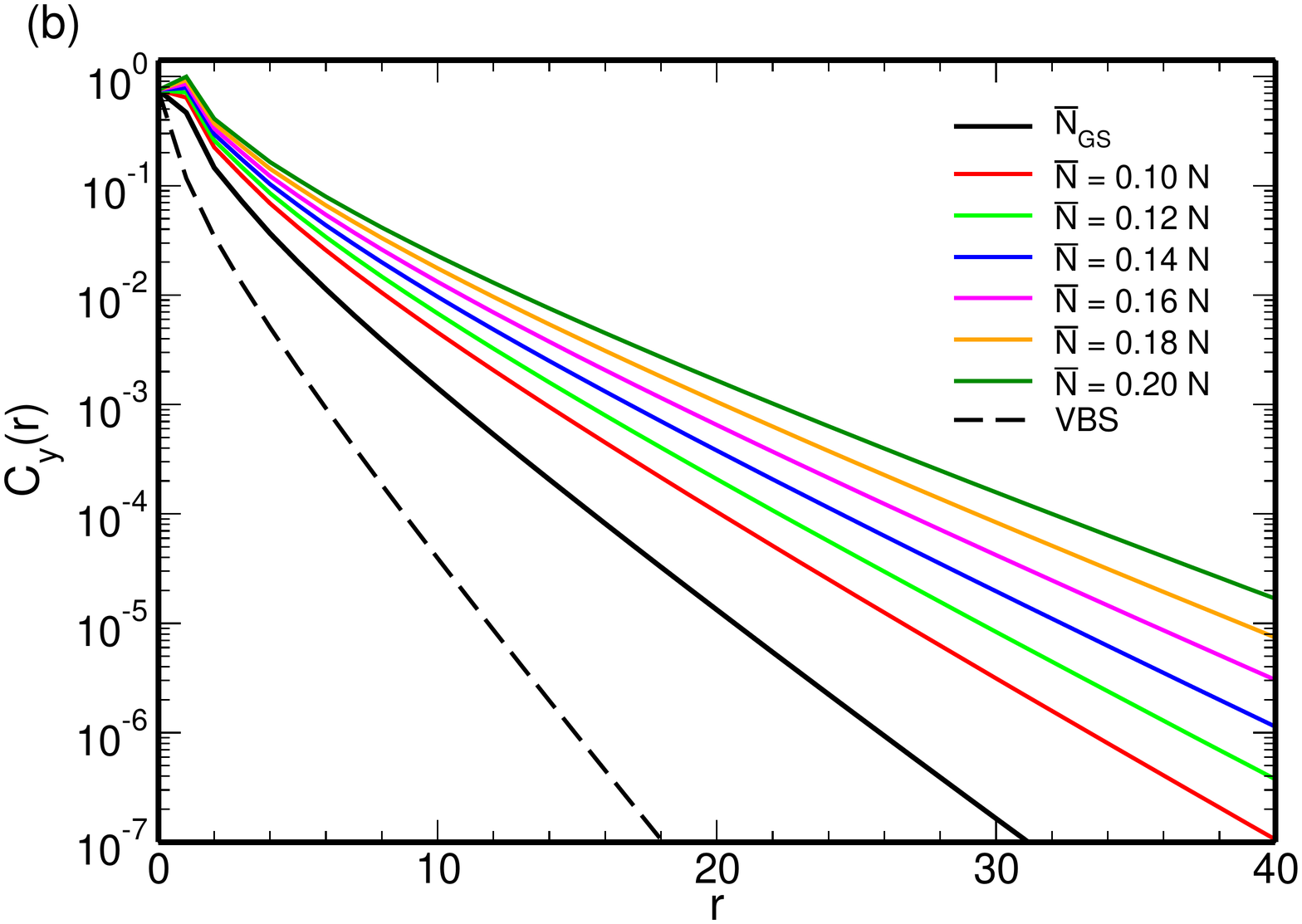}
}
\caption{Spin-spin correlation functions (odd distances $r$)  
  (a) $C_x(r)$ [Eq.~\eqref{cor-spin-x}] and
  (b) $C_y(r)$ [Eq.~\eqref{cor-spin-y}] of the many-triplon state \eqref{mf-wf}
  for $J_2 = 0.52\, J_1$. Mean-field results for different values 
  of the triplon number $\bar{N}$ (solid lines) are shown:
  $\bar{N} = \bar{N}_{GS}$ (black),
  $\bar{N} = 0.10\, N$ (red),
  $\bar{N} = 0.12\, N$ (green),
  $\bar{N} = 0.14\, N$ (blue),
  $\bar{N} = 0.16\, N$ (magenta),
  $\bar{N} = 0.18\, N$ (orange), and
  $\bar{N} = 0.20\, N$ (dark green).
  The corresponding harmonic results for the columnar VBS ground-state $|{\rm VBS}\rangle$ 
  are also included (dashed black line). }
\label{fig:cor-spin02}
\end{figure*}

\section{Correlation functions}
\label{sec:corr}

To further characterize the many-triplon states \eqref{mf-wf}, we calculate
spin-spin and dimer-dimer correlation functions and dimer order parameters.
We concentrate on two model configurations, 
$J_2 = 0.48 J_1$ and $J_2 = 0.52 J_1$,
since they exemplified the two distinct regions identified in Sec.~\ref{sec:mf} 
($J_2 \le 0.51 J_1$ and $J_2 > 0.51 J_1$) and
they are deep in the quantum paramagnetic region of the $J_1$-$J_2$ model \eqref{ham-j1j2}
(see Sec.~\ref{sec:model}), where the harmonic results are more reliable.
In addition to the (lowest-energy) many-triplon state \eqref{mf-wf} with
$\bar{N} = \bar{N}_{GS}$, we also consider states with $\bar{N} > \bar{N}_{GS}$. 
Moreover, comparisons with the corresponding harmonic results for the
columnar VBS state $|{\rm VBS}\rangle$ are also made.

\subsection{Spin--spin correlation functions}
\label{sec:corr-spin}

The spin-spin correlation functions $C_\alpha(r)$ are defined as
\begin{equation}
     C_\alpha(r) = \langle \bS_i \cdot \bS_{i + r\hat{\alpha}} \rangle,
\label{cor-spin}
\end{equation}
where $\bS_i$ is a spin-$1/2$ operator at the site $i$ of the 
{\sl original} square lattice and $\hat{\alpha} = \hat{x}, \hat{y}$ 
(recall that we set the lattice spacing of the original square
lattice $a = 1$).  
In terms of the spin operators $\bS^1_i$ and $\bS^2_i$ 
of the dimerized lattice $\mathcal{D}$ [see Fig.~\ref{fig:model}(a)],  
the spin-spin correlation functions \eqref{cor-spin} assume the form   
\begin{equation}
 C_x(r) = \left\{\begin{array}{ll}
               \langle \left(  \bS^1_i  \right)^2 \rangle, & r = 0, \\& \\       
               \langle \bS^1_i \cdot \bS^2_i \rangle,    & r = 1, \\&\\             
               \langle \bS^1_i \cdot \bS^1_j \rangle,    & r = | \bR_j - \bR_i | \ge  2, \\&\\
               \langle \bS^1_i \cdot \bS^2_j \rangle,    & r = | \bR_j - \bR_i | + 1 \ge 3, 
                        \end{array}\right.
\label{cor-spin-x}
\end{equation}
with $ \bR_j - \bR_i  = 2(j  - i)\hat{x}$ and
$\bR_i$ being a vector of the {\sl dimerized} lattice $\mathcal{D}$, and
\begin{equation}
 C_y(r) = \langle \bS^1_i \cdot \bS^1_j \rangle, \;\;\;\; r = | \bR_j - \bR_i |,
\label{cor-spin-y}
\end{equation}
with $ \bR_j - \bR_i = (j - i)\hat{y} $.

\begin{figure*}[t]
\centerline{\includegraphics[width=8.5cm]{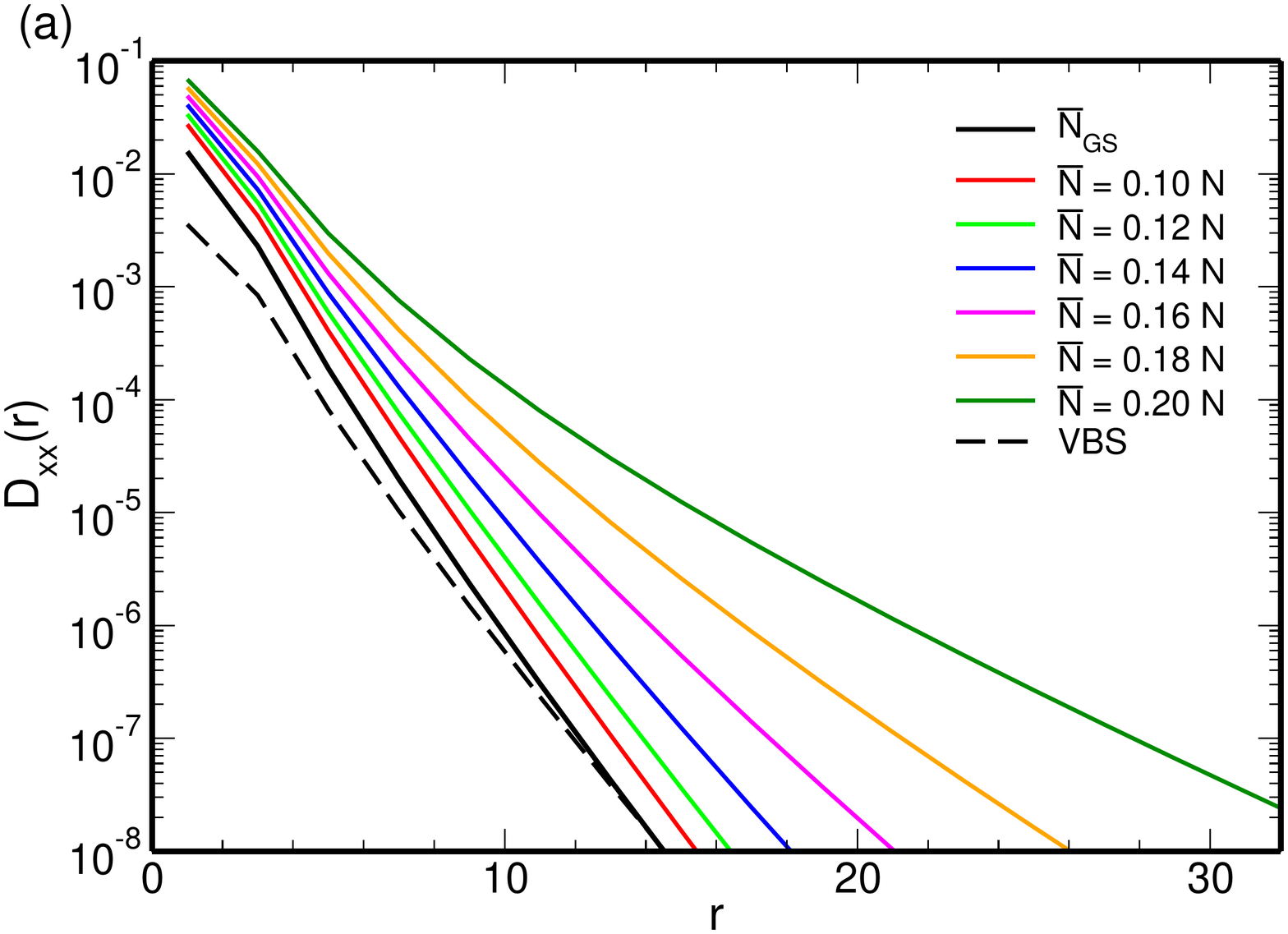}
                   \hskip0.5cm
                  \includegraphics[width=8.5cm]{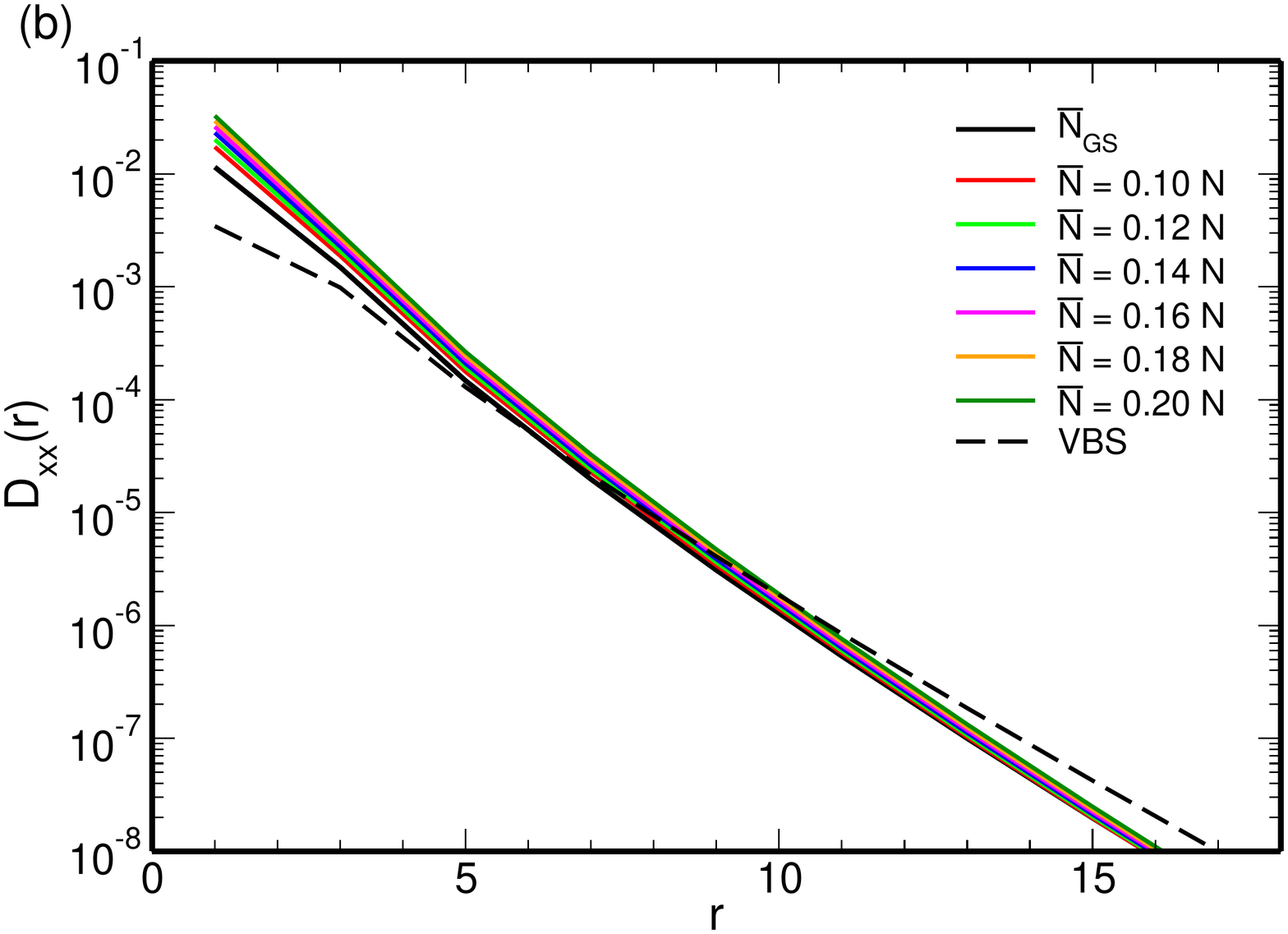}
}
\caption{Dimer-dimer correlation function (odd distances $r$)  
  $D_{xx}(r)$ [Eq.~\eqref{cor-dimer-x-1}] of the many-triplon state \eqref{mf-wf}
  for (a) $J_2 = 0.48\, J_1$ and (b) $J_2 = 0.52\, J_1$. Mean-field results for different values 
  of the triplon number $\bar{N}$ (solid lines) are shown:
  $\bar{N} = \bar{N}_{GS}$ (black),
  $\bar{N} = 0.10\, N$ (red),
  $\bar{N} = 0.12\, N$ (green),
  $\bar{N} = 0.14\, N$ (blue),
  $\bar{N} = 0.16\, N$ (magenta),
  $\bar{N} = 0.18\, N$ (orange), and
  $\bar{N} = 0.20\, N$ (dark green).
  The corresponding harmonic results for the columnar VBS ground-state $|{\rm VBS}\rangle$ 
  are also included (dashed black line).}
\label{fig:cor-dimer}
\end{figure*}

It is possible to show that 
\begin{eqnarray}
\langle \left(  \bS^1_i  \right)^2 \rangle &=& \frac{3}{4},
\;\;\;\;\;\;\;\;\;\;\;
\langle \bS^1_i \cdot \bS^2_i \rangle = -\frac{3}{4}\left( N_0 - I_{1,ii} \right),
\nonumber\\
&& \nonumber\\
\langle \bS^1_i \cdot \bS^1_j \rangle &=& 
  \frac{3}{2}\left[ | I_{1,ij} |^2 - | I_{2,ij} |^2 + N_0\left( I_{1,ij} + I_{2,ij}   \right) \right],
\nonumber \\
&& \nonumber \\
\langle \bS^1_i \cdot \bS^2_j \rangle &=& 
  \frac{3}{2}\left[ | I_{1,ij} |^2 - | I_{2,ij} |^2 - N_0\left( I_{1,ij} + I_{2,ij}   \right) \right],
\nonumber \\
\label{spin-correlations} 
\end{eqnarray}
with $i \not= j$. 
Here, the parameter $N_0$ is determined within the harmonic
approximation for the columnar VBS state $|{\rm VBS}\rangle$ 
[Eq.~\eqref{self-mu-nzero}], as already mentioned in Sec.~\ref{sec:mf}.
The integrals $I_{1,ij}$ and $I_{2,ij}$ are given by
\begin{eqnarray}
  I_{1,ij} &=& \frac{1}{N'}\sum_\bk \cos\left[ \bk\cdot\left( \bR_i - \bR_j \right)\right]f(\bk),
\nonumber \\
  I_{2,ij} &=& \frac{1}{N'}\sum_\bk \cos\left[ \bk\cdot\left( \bR_i - \bR_j \right)\right]\bar{f}(\bk),
\label{integrals}
\end{eqnarray}
with $N' = N/2$,
and the functions $f(\bk)$ and $\bar{f}(\bk)$ being defined as
\begin{equation}
 f(\bk) \equiv \langle t^\dagger_{\bk\alpha} t_{\bk\alpha} \rangle,
 \;\;\;\;\;\;\;
 \bar{f}(\bk) \equiv \langle t^\dagger_{\bk\alpha} t^\dagger_{-\bk\alpha} \rangle.
\end{equation}
With the aid of the Bogoliubov transformation \eqref{bogo-transf}, one
shows that, for the columnar VBS state $|{\rm VBS}\rangle$ within the
harmonic approximation, 
the functions $f(\bk)$ and $\bar{f}(\bk)$ read
\begin{eqnarray}
 f(\bk) &=& v^2_\bk
             = \frac{1}{2} \left( -1 + \frac{ A_\bk }{ \Omega_\bk } \right),
\nonumber \\
\bar{f}(\bk) &=& - v_\bk u_\bk 
                     =  - \frac{ B_\bk }{ 2\Omega_\bk },
\end{eqnarray}
with $u_\bk$ and $v_\bk$ being the Bogoliubov coefficients \eqref{bogo-coef}.
Similarly, using both Bogoliubov transformations \eqref{bogo-transf}
and \eqref{bogo-transf2}, one finds that, for the many-triplon state~\eqref{mf-wf},
\begin{eqnarray}
 f(\bk) &=& \left( u_\bk \bar{v}_\bk + v_\bk\bar{u}_\bk \right)^2 
\nonumber \\
&& \nonumber \\
      &=& \frac{1}{2\omega_\bk\bar{\Omega}_\bk}  
               \left( A_\bk\bar{A}_\bk  + B_\bk\bar{B}_\bk - \omega_\bk\bar{\Omega}_\bk \right),
\\
&& \nonumber \\
\bar{f}(\bk) &=& -\left( u_\bk \bar{v}_\bk + v_\bk\bar{u}_\bk \right)
                 \left( u_\bk \bar{u}_\bk + v_\bk\bar{v}_\bk \right)  
\nonumber \\
&& \nonumber \\
      &=& -\frac{1}{2\omega_\bk\bar{\Omega}_\bk}  
                \left( A_\bk\bar{B}_\bk  + B_\bk\bar{A}_\bk \right),
\label{fbar}
\end{eqnarray}
with $\bar{u}_\bk$ and $\bar{v}_\bk$ being the Bogoliubov coefficients
\eqref{bogo-coef-bcs}.

Figure~\ref{fig:cor-spin01} shows, for $J_2 = 0.48 J_1$, the spin-spin 
$C_x(r)$ [Fig.~\ref{fig:cor-spin01}(a)] and 
$C_y(r)$ [Fig.~\ref{fig:cor-spin01}(b)] correlation functions 
of the columnar VBS ground state (dashed black lines) 
and of the many-triplon state \eqref{mf-wf} (solid lines) with
different values of the triplon number $\bar{N}$. 
One notices that, for the columnar VBS state, both spin-spin
correlation functions decay exponentially, as expected for
a phase with a finite (triplet) excitation gap. Moreover, the
correlation length associated with $C_x(r)$ is larger than the one
related to $C_y(r)$. Such distinct behaviours found for the $C_x(r)$
and the $C_y(r)$ correlation functions are related to the symmetries of the
columnar VBS state: recall that we consider, in particular, a columnar
VBS state with dimers along the $x$ direction.
Similarly, for the many-triplon state, the two spin-spin correlation
functions also decay exponentially, regardless the triplon number
$\bar{N}$. 
For the columnar VBS ground state and the many-triplon states with 
$\bar{N} \le 0.12\, N$, the correlation lengths associated with the
correlation function $C_x(r)$ are approximately equal while, for
$\bar{N} \ge 0.14\, N$, the correlation length increases with
$\bar{N}$. Such features might be 
related to the fact that the excitation gap above the columnar VBS
ground state and the ones above the many-triplon state with 
$\bar{N}_{GS} \le \bar{N} \le  0.10\, N$ are approximately equal [see
Fig.~\ref{fig:disp}(a)]  while, for $\bar{N} > 0.10\, N$, the
excitation gap above the many-triplon state decreases as $\bar{N}$
increases. 
On the other hand, for $\bar{N} < 0.14\, N$, the correlation length
associated with the correlation function $C_y(r)$ seems to be less
sensitive to the excitation gap,
since it always increases with the triplon number $\bar{N}$. 
Again, these different features displayed by the $C_x(r)$ and the $C_y(r)$
correlation functions of the many-triplon state 
with $\bar{N} < 0.14\, N$ might be due to the
symmetries of the columnar VBS (reference) state.
Interesting, for larger values of the triplon number $\bar{N}$, the
behaviour of the $C_x(r)$ and the $C_y(r)$ correlation functions
are quite similar, indicating that, in this case, the many-triplon states
should display a more homogeneous singlet parttern
than the columnar VBS ground state.

\begin{figure*}[t]
\centerline{\hskip0.5cm
                  \includegraphics[width=3.5cm]{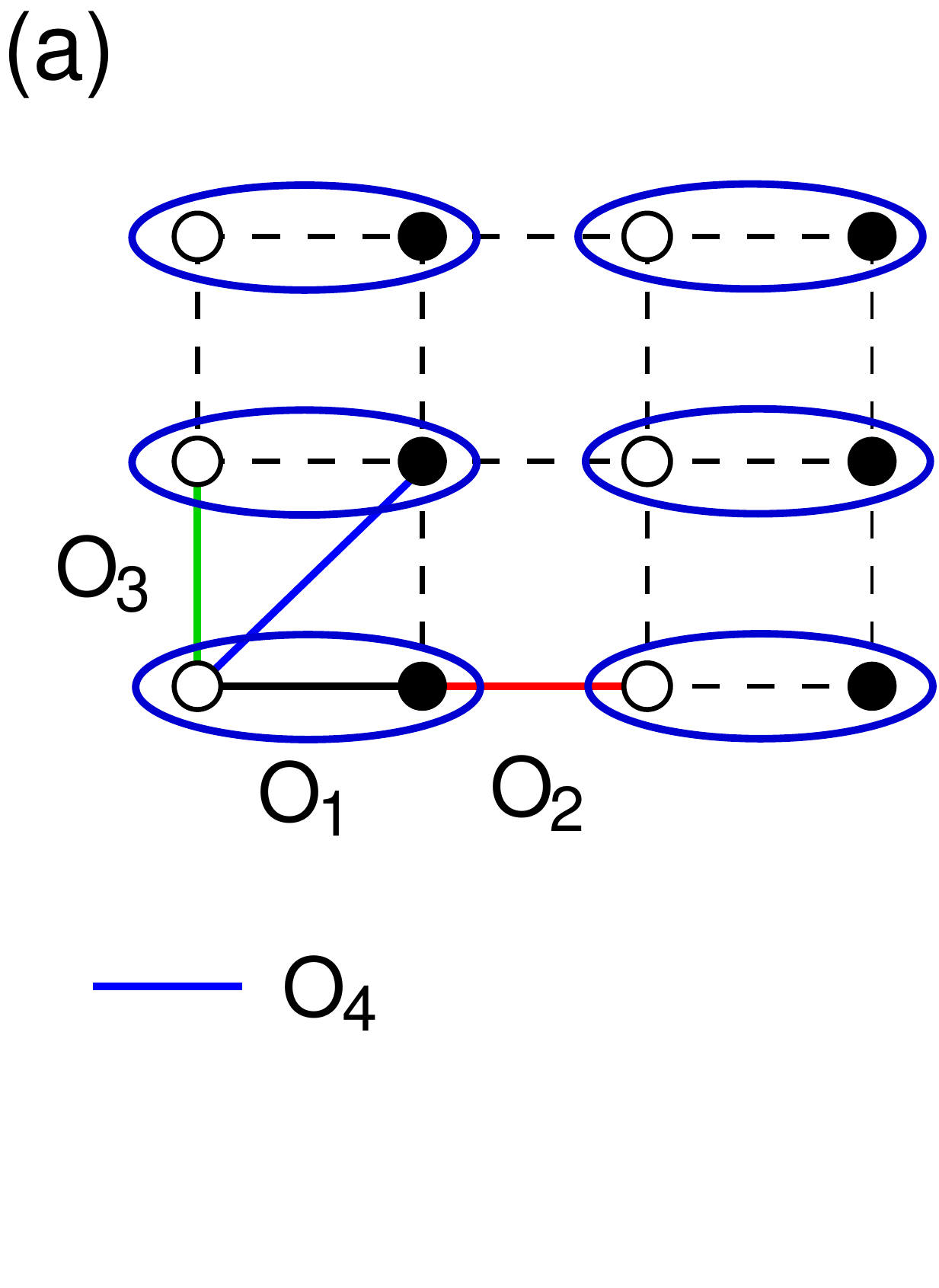}
                  \hskip0.7cm
                  \includegraphics[width=7.0cm]{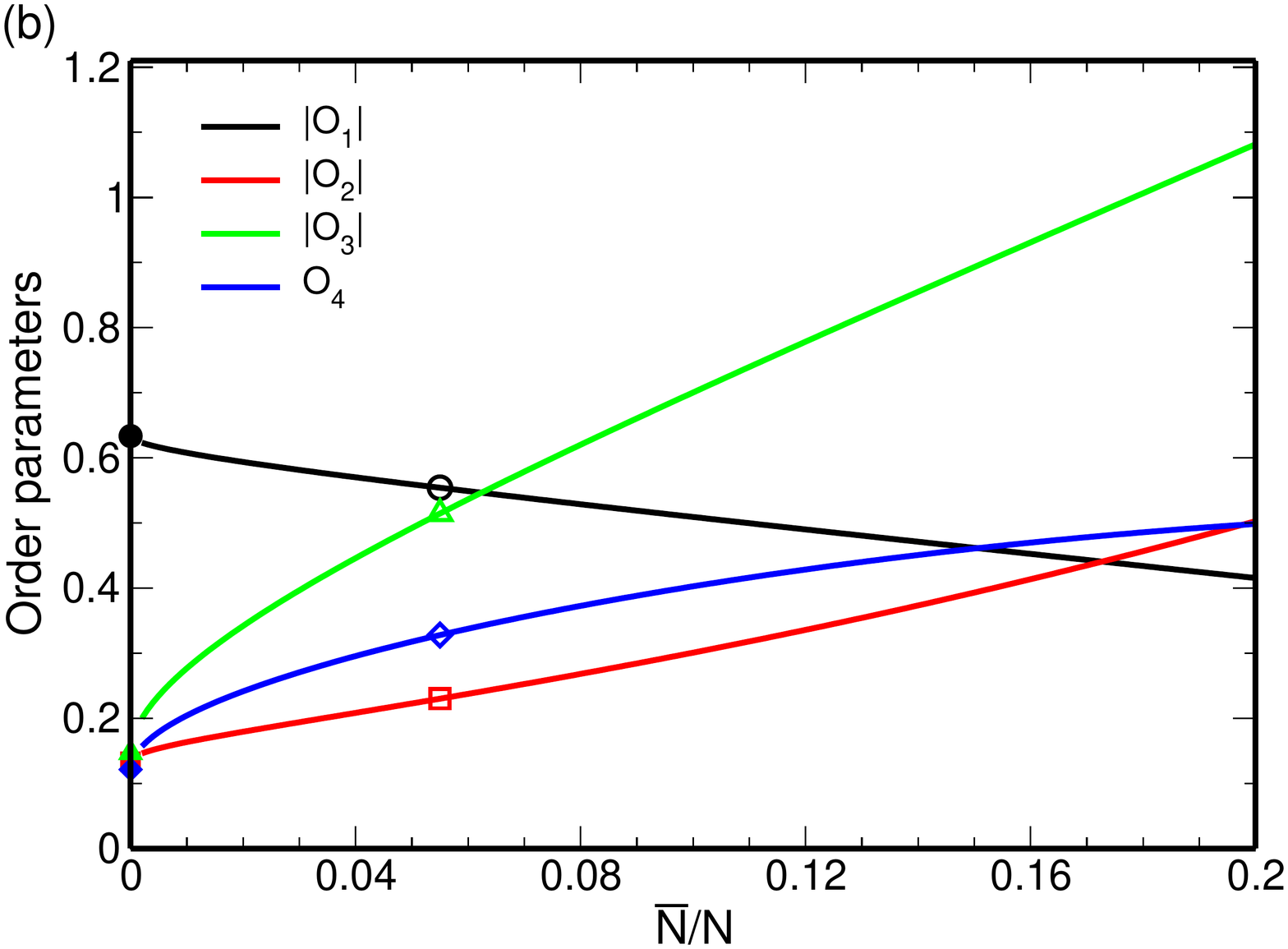}
                  \hskip0.2cm
                  \includegraphics[width=7.0cm]{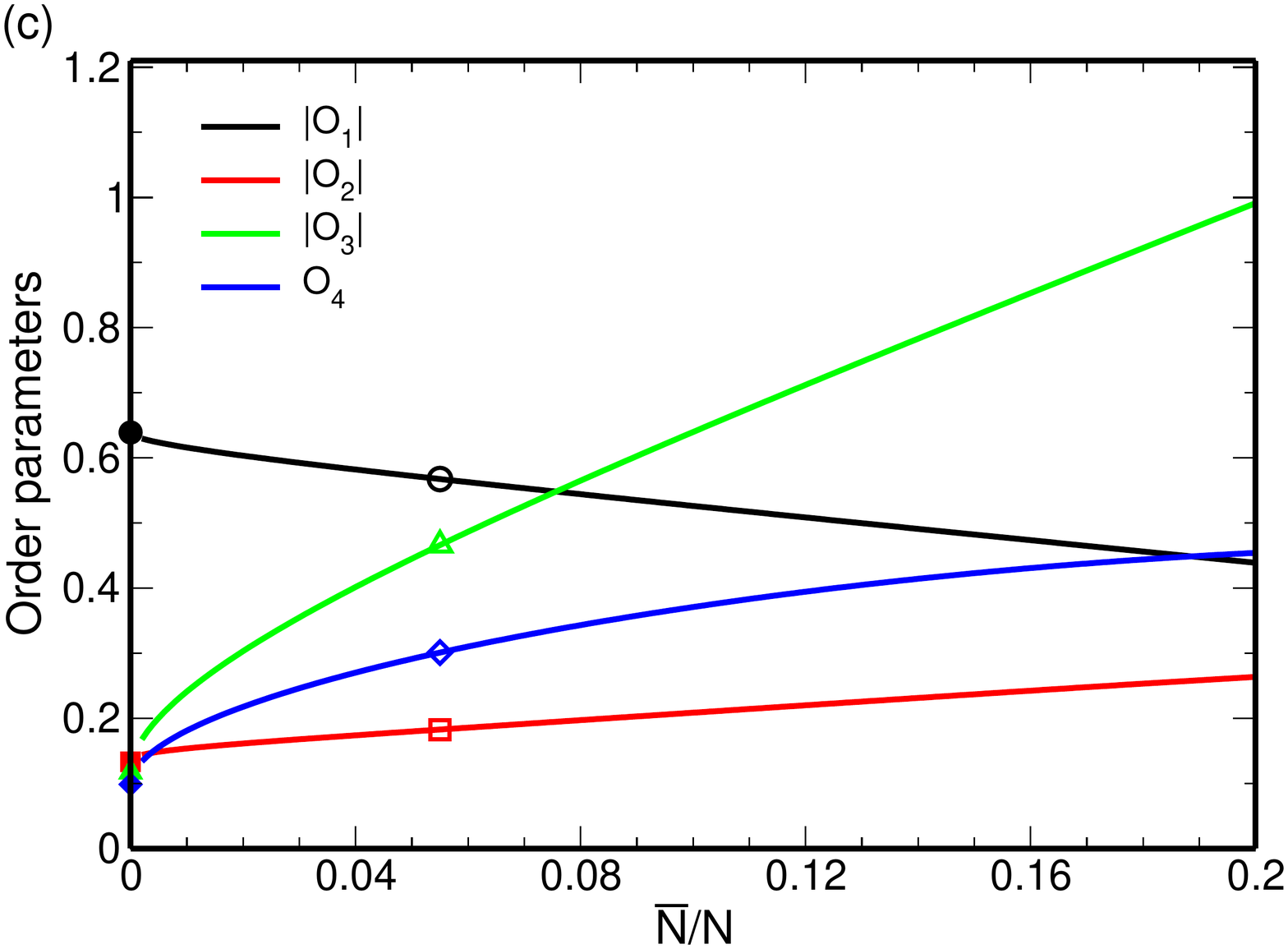}   }
\caption{(a) Schematic representation of the dimer order parameters 
  \eqref{dimer-orderpar} for the square lattice. 
  Dimer order parameters \eqref{dimer-orderpar} of the many-triplon 
  state \eqref{mf-wf} as a function of the ratio $\bar{N}/N$ for 
  (b) $J_2 = 0.48\, J_1$ and (c) $J_2 = 0.52\, J_1$.
  The solid symbols are the harmonic results for the columnar VBS
  ground state while the open symbols are the mean-field results for 
  the (lowest-energy) many-triplon state \eqref{mf-wf} with $\bar{N} = \bar{N}_{GS}$. 
}
\label{fig:orderpar}
\end{figure*}

The spin-spin $C_x(r)$ and $C_y(r)$ correlation functions of the 
columnar VBS ground state and the many-triplon state \eqref{mf-wf} 
with different values of the triplon number $\bar{N}$ for $J_2 = 0.52 J_1$
are shown in Figs.~\ref{fig:cor-spin02}(a) and (b), respectively. 
Similar to the configuration $J_2 = 0.48 J_1$, both correlation
functions exponentially decay with the distance $r$.
Interesting, for the many-triplon state, the correlation lengths
associated with the correlation function $C_x(r)$ are independent of
the triplon number $\bar{N}$ and they are almost equal to the corresponding 
one of the columnar VBS ground state. Again, this feature might be related to
the fact that the excitation gap above the many-triplon state slowly
decreases with the triplon number $\bar{N}$ and they are close to the
excitation gap above the columnar VBS ground state [see Fig.~\ref{fig:disp}(b)].
On the other hand, the correlation length associated with the 
correlation function $C_y(r)$ increases with $\bar{N}$, similar to the
behaviour found for  $J_2 = 0.48 J_1$.
Differently from the configuration $J_2 = 0.48 J_1$, here the 
behaviour of the $C_x(r)$ and the $C_y(r)$ correlation functions do not indicate that 
the many-triplon states with large $\bar{N}$ are constituted by a more
homogeneous singlet pattern than the corresponding columnar VBS state.

\subsection{Dimer-dimer correlation functions}
\label{sec:corr-dimer}

The dimer-dimer correlation functions $D_{\alpha\beta}(i,j)$ are defined as
\begin{equation}
   D_{\alpha\beta}( i, j ) = \langle B_\alpha( i ) B_\beta( j ) \rangle
                                 - \langle B_\alpha( i ) \rangle \langle B_\beta( j ) \rangle,
\label{cor-dimer}
\end{equation}
where the dimer operator $B_\alpha(i)$ reads
\begin{equation}
   B_\alpha(i) = \bS_i \cdot \bS_{i + \hat{\alpha}},
\label{aux-cor-dimer}
\end{equation}
with $\bS_i$ being a spin-$1/2$ operator at the site $i$ of the original
square lattice and $\hat{\alpha} = \hat{x}, \hat{y}$.
Similar to the spin-spin correlation functions \eqref{cor-spin},
we rewrite the dimer-dimer correlation functions \eqref{cor-dimer}
in terms of the spin operators $\bS^1_i$ and $\bS^2_i$ 
of the dimerized lattice $\mathcal{D}$. 
In particular, the dimer-dimer correlation function $D_{xx}(r)$
assumes the form   
\begin{eqnarray}
 D_{xx}(r) &=&              
        \langle \left( \bS^1_i\cdot\bS^2_i  \right)  \left( \bS^2_l\cdot\bS^1_j  \right)  \rangle 
\nonumber \\
   &-&   \langle\left( \bS^1_i\cdot\bS^2_i  \right) \rangle \langle \left( \bS^2_l\cdot\bS^1_j \right)  \rangle,      
\label{cor-dimer-x-1}
\end{eqnarray}
for $r = |\bR_j - \bR_i| - 1$, and 
\begin{eqnarray}
 D_{xx}(r) &=&              
       \langle \left( \bS^1_i\cdot\bS^2_i  \right)  \left( \bS^1_j\cdot\bS^2_j  \right)  \rangle 
\nonumber \\ 
   &-&  \langle\left( \bS^1_i\cdot\bS^2_i  \right) \rangle \langle \left( \bS^1_j\cdot\bS^2_j \right)  \rangle,    
\label{cor-dimer-x-2}
\end{eqnarray} 
for $r = |\bR_j - \bR_i|$, with      
$\bR_j - \bR_i = 2( j - i )\hat{x}$,
$\bR_j - \bR_l = 2\hat{x}$, and
$\bR_i$ being a vector of the dimerized lattice $\mathcal{D}$. 
Here, it is also possible to express $D_{xx}(r)$ in terms of the integrals
\eqref{integrals}:
\begin{eqnarray}
 D_{xx}(r) &=&  -\frac{3}{8}N_0\left( I_{1,ij} + I_{2,ij}  \right) \left( I_{1,il} + I_{2,il} \right)
\nonumber \\
 &&  + \frac{3}{4} I_{1,jl} \left( I_{1,ij}I_{1,li}  + I_{2,ij}I_{2,li} \right) 
\nonumber \\
  &&  - \frac{3}{4} I_{2,jl} \left( I_{1,ij}I_{2,li}  + I_{2,ij}I_{1,li}\right),                            
\label{cor-dimer-x-3}
\end{eqnarray}
for $r = |\bR_j - \bR_i| - 1$, and 
\begin{eqnarray}
 D_{xx}(r) &=& \frac{3}{16}\left[  | I_{1,ij} |^2 + | I_{2,ij} |^2  \right],                    
\label{cor-dimer-x-4}
\end{eqnarray} 
for $r = |\bR_j - \bR_i|$, with the parameter $N_0$ being determined
within the harmonic approximation for the columnar VBS state, see
Eq.~\eqref{self-mu-nzero}. 
Therefore, with the aid of Eqs.~\eqref{integrals}--\eqref{fbar}, we
can determine the dimer-dimer correlation function $D_{xx}(r)$  
of the columnar VBS state and the
many-triplon state \eqref{mf-wf}.

The dimer-dimer correlation function $D_{xx}(r)$ of the columnar VBS
ground state and of the many-triplon state \eqref{mf-wf}  with different
values of the triplon number $\bar{N}$ for $J_2 = 0.48 J_1$ are
shown in Fig.~\ref{fig:cor-dimer}(a).  
Similar to the columnar VBS state, the dimer-dimer
correlation function of the many-triplon states decay exponentially, 
regardless the value of the triplon number $\bar{N}$. 
Such a behaviour indicates that the singlet excitation gap above the
many-triplon state is finite.
Moreover, the correlation length associated with $D_{xx}(r)$ increases
with the triplon number $\bar{N}$.
These features are quite similar to the ones found for the spin-spin
correlation function $C_x(r)$ [Fig.~\ref{fig:cor-spin01}(a)], although
the dimer correlation decays faster than the corresponding spin one
and the dimer correlation length always increases with the triplon
number $\bar{N}$, even for $\bar{N} \le 0.12\, N$.

For $J_2 = 0.52 J_1$, we also found that the correlation
function $D_{xx}(r)$ of both columnar VBS ground state and many-triplon
states decay exponentially, see Fig.~\eqref{fig:cor-dimer}(b). 
Differently from the configuration $J_2 = 0.48 J_1$, here the
correlation length associated with $D_{xx}(r)$ is independent of
the triplon number $\bar{N}$ and it is close to the correlation length 
of the columnar VBS state. 
Again, these features are quite similar to the ones found for the
corresponding spin-spin correlation function $C_x(r)$
[Fig.~\ref{fig:cor-spin02}(a)], apart from the fact that the dimer
correlations decay faster than the corresponding spin ones.

\begin{figure*}[t]
\centerline{\includegraphics[width=8.5cm]{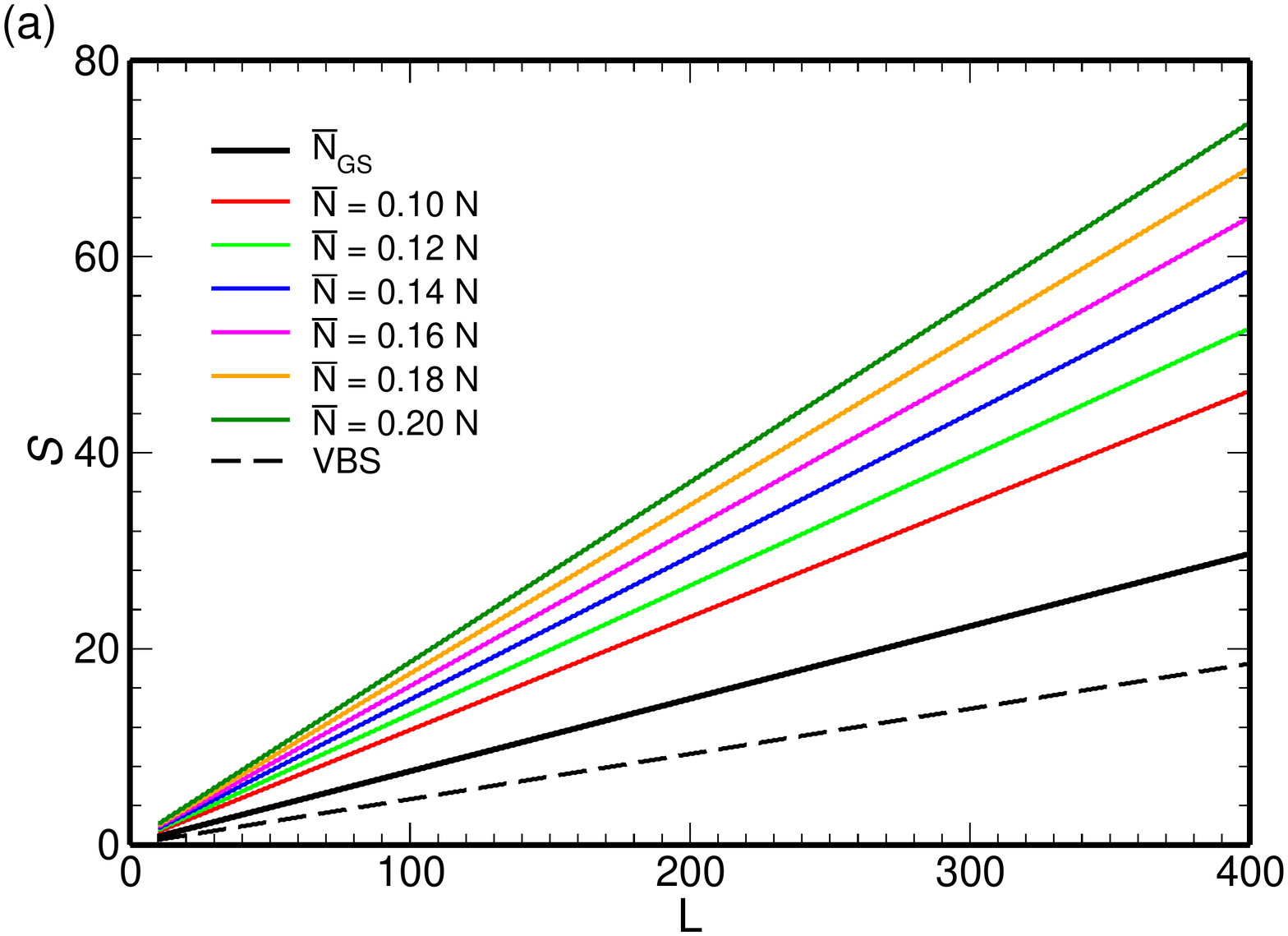}
                   \hskip0.5cm
                   \includegraphics[width=8.5cm]{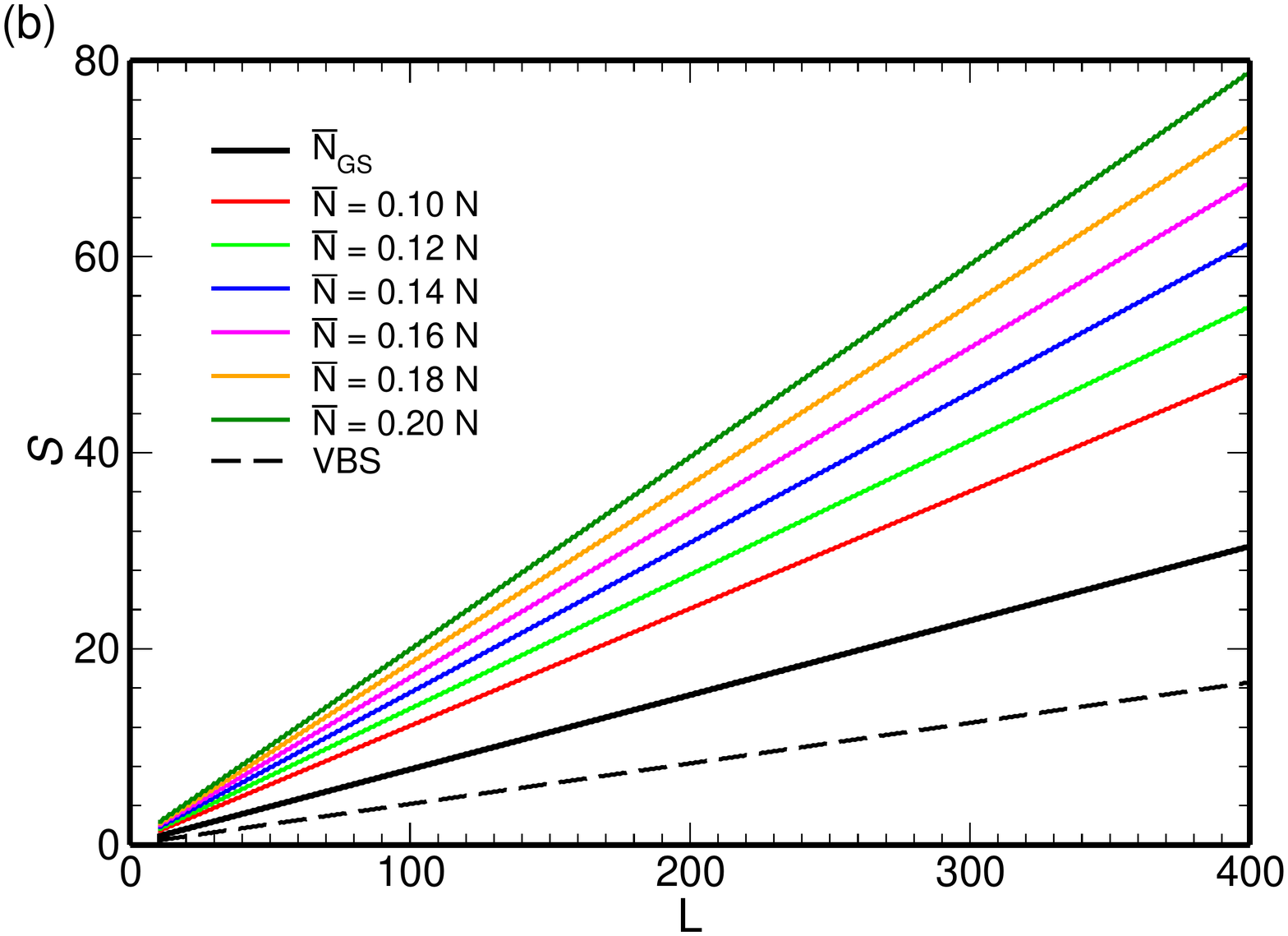}
}
\caption{The von Neumann entanglement entropy $\mathcal{S}$ as a
  function of the size $L$ of the (line) subsystem $A$ of the
  many-triplon state \eqref{mf-wf}
  for (a) $J_2 = 0.48\, J_1$ and (b) $J_2 = 0.52\, J_1$.
  Results for different values 
  of the triplon number $\bar{N}$ (solid lines) are shown:
  $\bar{N} = \bar{N}_{GS}$ (black),
  $\bar{N} = 0.10\, N$ (red),
  $\bar{N} = 0.12\, N$ (green),
  $\bar{N} = 0.14\, N$ (blue),
  $\bar{N} = 0.16\, N$ (magenta),
  $\bar{N} = 0.18\, N$ (orange), and
  $\bar{N} = 0.20\, N$ (dark green).
  The corresponding harmonic results for the columnar VBS ground-state $|{\rm VBS}\rangle$ 
  are also included (dashed black line). 
}
\label{fig:entropy}
\end{figure*}

\subsection{Dimer order parameters}
\label{sec:dimer-order}

Let us now consider the following dimer order parameters:
\begin{eqnarray}
O_1 &=&  \langle \bS^1_i \cdot \bS^2_i \rangle,  
\nonumber \\
&& \nonumber \\
O_2 &=&  \langle \bS^2_i \cdot \bS^1_j \rangle,  
 \;\;\;\;\;\;\; \bR_j - \bR_i = 2a\hat{x},
\nonumber \\
&& \nonumber \\
O_3 &=&  \langle \bS^1_i \cdot \bS^1_j \rangle,  
 \;\;\;\;\;\;\; \bR_j - \bR_i = a\hat{y},
\nonumber \\
&& \nonumber \\
O_4 &=&  \langle \bS^1_i \cdot \bS^2_j \rangle,  
 \;\;\;\;\;\;\; \bR_j - \bR_i = a\hat{y},
\label{dimer-orderpar}
\end{eqnarray}
where $\bS^1_i$ and $\bS^2_i$ are spin operators at the site $i$ of the
dimerized lattice $\mathcal{D}$ and $\bR_i$ is the vector associated with
the site $i$. The four dimer order parameters $O_i$ are illustrated in
Fig.~\ref{fig:orderpar}(a).  
With the aid of Eq.~\eqref{spin-correlations}, one sees that the dimer
order parameters \eqref{dimer-orderpar} can be written in terms of the
integrals $I_{1,ij}$ and $I_{2,ij}$ [Eq.~\eqref{integrals}], and
therefore, they can be easily calculated for both columnar VBS state and
many-triplon state \eqref{mf-wf}.

Figure \ref{fig:orderpar} shows the behaviour of the dimer
order parameters \eqref{dimer-orderpar} as a function of the triplon
number $\bar{N}$ for $J_2 = 0.48\, J_1$ [Fig.~\ref{fig:orderpar}(b)] 
and $J_2 = 0.52\, J_1$ [Fig.~\ref{fig:orderpar}(c)]. 
The results for the columnar VBS ground state are indicated by
the solid symbols, while the results for the (lowest-energy) many-triplon
state with $\bar{N} = \bar{N}_{GS}$ are indicated by the open ones. 
As expected, for the columnar VBS ground state, we found that 
$| O_1 | > | O_2 | \approx | O_3| \approx | O_4|$ while,
for the lowest-energy many-triplon state with $\bar{N} = \bar{N}_{GS}$, we have 
$| O_1 | \approx | O_3 | > | O_2| \approx | O_4|$, i.e.,
apart from the value of the dimer order parameter $O_3$ 
(see discussion below),  such a state might display the same features
of the columnar VBS state.   
For a large triplon number $\bar{N}$, in particular, $\bar{N} \sim 0.17$,
we notice that,  
$| O_1 | \approx | O_2 | \approx |O_4| \sim 0.46$ for $J_2 = 0.48\, J_1$,  
while
$| O_1 | \approx |O_4| \sim 0.45 > |O_2|$ for $J_2 = 0.52 J_1$. 
Therefore, the dimer order parameters indicate that the many-triplon
states with large $\bar{N}$ might display a more homogeneous singlet
pattern for $J_2 = 0.48\, J_1$ than for $J_2 = 0.52\, J_1$.
Recall that such features are in agreement with the ones found for
the spin-spin $C_x(r)$ and $C_y(x)$ correlation functions, see
Sec.~\ref{sec:corr-spin}.

Finally, concerning the behaviour of the dimer order parameter $O_3$
with the triplon number $\bar{N}$, it is not clear, at the moment, the
reason $O_3$ increases so fast with $\bar{N}$. We believe it could be an
artefact of the approximations involved in our mean-field calculations. Indeed,
such an artefact could also affect the behaviour of the spin-spin
correlation function $C_y(r)$: recall that the correlation length associated
with $C_y(r)$ seems to be less sensitive to the excitation
gap than the correlation length related to the correlation
function $C_x(r)$, see Sec.~\ref{sec:corr-spin}.

\section{Entanglement entropy}
\label{sec:entropy}

In this section, we calculate the bipartite von Neumann entanglement
entropy $\mathcal{S}$ of the columnar VBS and the many-triplon states. In
particular, we follow the procedure \cite{leite19}, that was recently 
employed to calculate the entanglement entropies of the ground state
of spin-$1/2$ dimerized Heisenberg AFMs on a square lattice.

The bipartite entanglement entropy of the ground state $|\Psi\rangle$
of a system $S$ is defined, for instance, as the von Neumann entropy
\cite{grover13,review-nicolas},    
\begin{equation}
  \mathcal{S} =\mathcal{S}(\rho_A) = - {\rm Tr}\left( \rho_A \ln \rho_A \right),
\label{neumann}  
\end{equation}
where $A$ is a subsystem (arbitrary size and shape), 
$\bar{A}$ is its complementary such that the system $S = A \cup \bar{A}$,
and $\rho_A = {\rm Tr}_{\bar{A}}  |\Psi\rangle \langle \Psi |$ is the
reduced density matrix of the subsystem $A$.  
For a generic quadratic Hamiltonian written in terms of boson
operators, it is possible to show that the entanglement entropy
\eqref{neumann} assumes the form
\begin{equation}
  \mathcal{S} = \sum_{m = 1}^{N_A}  \sum_{\epsilon = \pm 1} 
        \epsilon\left( \frac{\mu_m + \epsilon }{2} \right)
                     \ln\left( \frac{\mu_m + \epsilon }{2} \right),
\label{Smu-neumann}
\end{equation}
where $\mu^2_m$ are the eigenvalues of the so-called  
correlation matrix $C$ (for the definition, see Eq.~(41) of
Ref.~\cite{leite19}) and $N_A < N$ is the number of sites of the
subsystem $A$. 
In particular, for a one-dimensional (line) subsystem $A$ 
[a spin chain of size $L$, see Fig.~\ref{fig:model}(b)],
the eigenvalues $\mu^2_m$ of the correlation matrix can be
analytically calculated.      
For the columnar VBS ground state described by the harmonic
Hamiltonian \eqref{ham-harmonic}, we have
\begin{equation}
    \mu_{m}^2 =    
       \left( \frac{1}{N_y} \sum_{k_y} \frac{ A_{m,k_y} }{ \omega_{m,k_y} } \right)^2  
    - \left( \frac{1}{N_y} \sum_{k_y} \frac{ B_{m,k_y} }{\omega_{m,k_y} } \right)^2,                
\label{muAB}
\end{equation}
where the index $m = 1, 2, \cdots$ and $N_A$ 
is related to the momentum $k_x$ parallel to the
system-subsystem boundary, 
\begin{equation}   
   k_x = -\frac{\pi}{2} + \frac{2\pi (m - 1)}{L+2}, 
\label{kx-par}
\end{equation}
with $N' = N_AN_y$ and $N_A = (L + 2)/2$.
Moreover, $A_{\bk}$ [Eq.~\eqref{ak}] and $B_{\bk}$ [Eq.~\eqref{bk}] are the
coefficients of the harmonic Hamiltonian \eqref{ham-harmonic}, and
$\omega_\bk$ is the energy of the triplons \eqref{omega-harmonic}.   
Similarly, for the many-triplon state \eqref{mf-wf} described by the
mean-field Hamiltonian \eqref{h-mf}, the eigenvalues of the correlation
matrix are also given by Eq.~\eqref{muAB}, but with the replacements:
$A_\bk \rightarrow \bar{A}_\bk$,
$B_\bk \rightarrow \bar{B}_\bk$ [Eq.~\eqref{abbark}], and
$\omega_\bk \rightarrow \bar{\Omega}_\bk$ [Eq.~\eqref{omega-bcs}].

The bipartite von Neumann entanglement entropy $\mathcal{S}$ 
in terms of the subsystem size $L$ of the
columnar VBS ground state and the many-triplon state with different values of
the triplon number $\bar{N}$ for $J_2 = 0.48\, J_1$ and
$J_2 = 0.52\, J_1$ are shown in Figs.~\ref{fig:entropy}(a) and (b), respectively. 
As expected for a two-dimensional gapped phase \cite{rmp-area-law},
we find that the entanglement entropy is dominated by an area law for
both columnar VBS and many-triplon states: we fit the data shown in
Figs.~\ref{fig:entropy}(a) and (b) with the curve 
\begin{equation}
  \mathcal{S} = a L + b\ln L + c,
\label{fit-ee}
\end{equation}
and find that the coefficient $b < 10^{-5}$, see
Table~\ref{table:entropy} for details. 
Moreover, for a given subsystem size $L$, we notice that the
entanglement entropy increases as the triplon number $\bar{N}$
increases. Such a feature is similar to the one found for the square
lattice dimerized Heisenberg AFMs 
(see Figs.~6(a) and (b) of Ref.~\cite{leite19}): as the dimerization
decreases and the system approaches the N\'eel-VBS quantum phase
transition, the number of triplets $t$ increases, and therefore, the
entanglement entropy increases although it seems not to diverge at the
quantum critical point.

\begin{table}[b]
\centering
\caption{Coefficients $a$, $b$, and $c$ obtained by fitting the von
  Neumann entanglement entropies $\mathcal{S}$ shown in
  Figs.~\ref{fig:entropy}(a) and (b) with the curve \eqref{fit-ee}.} 
\begin{tabular}{lcccccccc}
\hline\hline
 &&  & $J_2  = 0.48\, J_1$  &  && &  $J_2 = 0.52\, J_1$ & \\ 
$\bar{N}$ && a & b & c && a & b & c \\ 
\hline 
   VBS &\quad \quad & 0.05   &  1.82e-05  &  0.05
           &\quad\quad &           0.04   &  1.18e-08  &   0.04 \\ 

 $\bar{N}_{GS}$  &  & 0.07  & 4.28e-05 & 0.12  &&  0.08 & 4.05e-05  & 0.12 \\ 

 0.10 &  & 0.11  & 7.51e-07 & 0.18  &&  0.12 & 2.14e-08  &  0.18 \\ 

 0.12 &  & 0.13  & 8.17e-09 & 0.21  &&  0.14 & 6.64e-09  & 0.21  \\ 

 0.14 &  & 0.15  & 5.93e-09 & 0.23  &&  0.15 & 5.16e-09  &  0.23\\ 

 0.16 &  & 0.16  & 5.03e-09 & 0.25  &&  0.17 & 4.69e-09  &  0.25\\ 

 0.18 &  & 0.17  & 5.02e-09 & 0.27  &&  0.18 & 5.19e-09 &  0.27\\ 

 0.20 &  & 0.18  & 4.79e-09 & 0.28  &&  0.19 & 4.49e-09  & 0.29 \\ 
\hline\hline
\end{tabular} 
\label{table:entropy}
\end{table}

\section{Summary and discussion}
\label{sec:summary}

The bond-operator representation for spin operators introduced by
Sachdev and Bhatt \cite{sachdev90} is an interesting formalism that
allows us to analytically describe a VBS phase of a Heisenberg model. 
Not only dimerized VBS phases (as the columnar VBS one discussed in
this paper) could be described within this formalism, but it could
also be employed to study VBS phases with larger unit cells, such as  
the tetramerized plaquette VBS \cite{doretto14,zhito96}.  
Indeed, the bond-operator formalism is quite suitable for the
description of a VBS phase: In this case, it is possible to identify a
singlet pattern (reference state) and label the different
spins that constituted each singlet (unit cell); 
although each spin within the unit cell has a distinct
representation in terms of the bosonic bond operators [see, e.g.,  Eq.~\eqref{spin-bondop}],  
the mapping from a spin Hamiltonian to an effective boson one is
well defined, since the singlets are regularly distributed in space.

It would be interesting to apply the bond-operator formalism to
describe the another set of quantum paramagnetic phases, the spin liquids. 
It would be an alternative to the Schwinger boson formalism
\cite{assa} that is usually employed to analytically study
spin-liquid phases \cite{yang16}.
However, such an application is rather difficult to implement:
in this case, it is not possible to define an initial singlet pattern
(reference state), and therefore, a mapping from a Heisenberg model to
an effective boson one is not well-defined.  
In contrast, in the Schwinger boson formalism, all spin operators have the same
expansion in terms of the boson operators, and therefore, a mapping
from a spin model to a boson one can be done without a reference state 
(the initial singlet pattern).   
As mentioned in Sec.~\ref{sec:intro}, one motivation to study a system
within a fixed number of triplons $b$ above a VBS ground state  
is to check whether the possible many-triplon state could restore the
lattice symmetries broken when the VBS phase sets in as well as to
check whether such a many-triplon state could describe a spin-liquid phase.
If so, then the formalism discussed here could be used to describe
(gapped) spin-liquid phases within the bond-operator representation
\cite{sachdev90}.
Important, while the Schwinger boson formalism is based on {\sl spinon}
degrees of freedom, the bond-operator one would be based on
{\sl spinon-pair} ones, i.e., the boson operators $a$ defined in Eq.~\eqref{bogo-transf2}.

For the square lattice spin-$1/2$ $J_1$-$J_2$ AFM Heisenberg model
with the columnar VBS as a reference state, our mean-field results
indicate that the many-triplon state \eqref{mf-wf} is stable,
although the lowest-energy one has a quite small number of
triplons $\bar{N} = \bar{N}_{GS}$ (see Fig.~\ref{fig:nbarmax}).   
Therefore, we would expect that the columnar VBS ground state and the many-triplon
state with $\bar{N}= \bar{N}_{GS}$ would have similar features. 
Indeed, for $J_2 = 0.48 J_1$ and $J_2 = 0.52 J_1$, we found that 
the spin-spin $C_x(r)$ [Figs.~\ref{fig:cor-spin01}(a) and \ref{fig:cor-spin02}(a)] 
and the dimer-dimer $D_{xx}(r)$ (Fig.~\ref{fig:cor-dimer}) correlation functions  
of both states decay exponentially with correlation lengths
approximately equal. Such features are related to the fact that
the excitation gap of both states are quite close (see Fig~\ref{fig:gap}).    
On the other hand, the correlation length associated with the spin-spin
correlation function $C_y(r)$ of the columnar VBS state is smaller
than the one of the lowest-energy many-triplon state
[see Figs.~\ref{fig:cor-spin01}(b) and \ref{fig:cor-spin02}(b)]. 
Moreover, although the excitation gap above the two states are
approximately equal, the corresponding excitation spectra are indeed
distinct,  as exemplified 
for $J_2 = 0.48 J_1$ [Fig.~\ref{fig:disp}(a)] 
and $J_2 = 0.52 J_1$ [Fig.~\ref{fig:disp}(b)].
In particular, the momenta associated with the excitation gap are
equal for both columnar VBS and lowest-energy many-triplon states
only for $J_2 \le 0.51 J_1$.

In addition to the lowest-energy many-triplon state with $\bar{N} = \bar{N}_{GS}$, 
we also study (high energy) many-triplon states with $\bar{N} > \bar{N}_{GS}$
for configurations deep in the disorder region of the model
\eqref{ham-j1j2}, where our mean-field results are more reliable.
For $J_2 = 0.48 J_1$ and $J_2 = 0.52 J_1$, we found that the
excitation gaps are finite, they decrease with the triplon number $\bar{N}$,
and they are located at the $Y$ ($J_2 = 0.48 J_1$) and $M$ ($J_2 = 0.48 J_1$) 
points of the first Brillouin zone (Fig.~\ref{fig:disp}).
Moreover, we also found that the spin-spin 
(Figs.~\ref{fig:cor-spin01} and \ref{fig:cor-spin02}) 
and the dimer-dimer (Fig.~\ref{fig:cor-dimer}) correlation functions
of the many-triplon states decay exponentially, 
regardless the triplon number $\bar{N}$. 
In fact, the behaviour of the spin-spin correlation functions
indicates that, only for $J_2 = 0.48 J_1$, the many-triplon states
with large triplon number $\bar{N}$ might display a more homogeneous
singlet pattern than the columnar VBS state.
Interesting, DMRG calculations also found distinct features for
the model parameters $J_2 = 0.48 J_1$ and $J_2 = 0.52 J_1$: 
Gong {\sl et al.} \cite{gong14} 
found evidences for a gapless phase for $0.44\, J_1 < J_2 < 0.50\, J_1$ 
and a plaquette VBS ground state for $0.50\, J_1 < J_2 < 0.61\, J_1$;  
the calculations of Wang and Sandvik \cite{wang18} indicated that  
a gapless spin-liquid phase sets in for $0.46\, J_1 < J_2 < 0.52\, J_1$ while
a (columnar) VBS ground state, for $0.52\, J_1 < J_2 < 0.62\, J_1$.
Although a proper comparison between our results and the DMRG ones is
rather difficult, our procedure seems to be able to distinguish the parameter
regions $J_2 \lesssim 0.51  J_1$ and $J_2 \gtrsim 0.51 J_1$ of the
square lattice $J_1$-$J_2$ model. 
One should also mention a quite recent variational calculation 
based on Gutzwiller-projected fermionic wave-functions \cite{ferrari20}
and results based on machine-learning methods \cite{nomura20} that
agree with the findings of Ref.~\cite{wang18}.

Monte Carlo simulations were employed to calculated the spin-spin and
the dimer-dimer correlation functions of the (nearest-neighbor)
resonating-valence-bond (RVB) state on the square lattice 
\cite{albuquerque10,tang11}. 
It was found that the spin correlations decay exponentially while the
dimer ones decay algebraically with an exponent $\alpha \sim 1.2$.
Such a behaviour is similar to the classical dimer model, although the
dimer correlations of the RVB state decay more slowly than the ones of
the classical dimer model ($\alpha = 2.0$). 
Interesting, Monte Carlo calculations for the nearest-neighbor RVB
state but on the triangular and kagome lattices \cite{julia12,yang12}
and for a RVB state on the square lattice whose longest valence bonds
are between next-nearest-neighbors \cite{yang12} found that both spin
and dimer correlations decay exponentially. 
Comparing with our mean-field results for the many-triplon state
\eqref{mf-wf}, one sees that it displays the same features of the
next-nearest-neighbor RVB state on the square lattice.
Since the spin correlations for $J_2 = 0.48 J_1$ also indicate that
the many-triplon states with large $\bar{N}$ may be characterized by a more
homogeneous singlet pattern than the columnar VBS ground state, we would expect
that, in this case, the many-triplon state could describe a spin-liquid phase.

To further characterize the many-triplon states with large
$\bar{N}$, we determined the behaviour of the dimer order parameters 
\eqref{dimer-orderpar} with the triplon number $\bar{N}$ (Fig.~\ref{fig:orderpar}). 
For $J_2 = 0.48 J_1$, we found that the dimer order parameters seem not to
converge to the same value as $\bar{N}$ increases, although
$| O_1 | \approx | O_2 | \approx |O_4| \sim 0.46$ when $\bar{N} \sim 0.17$.
Such a behaviour of the dimer order parameters 
indicates that the many-triplon states with large $\bar{N}$
do not correspond to a spin-liquid state: In a spin-liquid phase, the
dimer order parameters are approximately equal, as found, e.g., for
the nearest-neighbor RVB state on the kagome lattice in Ref.~\cite{julia12};
for the many-triplon state, we would expect that a transition from a
columnar VBS state to spin-liquid one as $\bar{N}$ increases may be
signaled by a convergence of the dimer order parameters to a single
value, i.e,  
$| O_1 | \sim | O_2 | \sim | O_3| \sim | O_4|$ for $\bar{N} >\bar{N}_c$,
a feature that is not observed.

Finally, we also calculated the bipartite von Neumann entanglement
entropy of the columnar VBS and many-triplon states (Fig.~\ref{fig:entropy}). 
Fitting the data with the curve \eqref{fit-ee}, we found that the
entanglement entropies obey an area law, as expected 
for a two-dimensional gapped phase, and that the coefficient $c > 0$
and it increases with the triplon number $\bar{N}$ for both 
$J_2 = 0.48 J_1$ and $J_2 = 0.52 J_1$. These results corroborate the
fact that the many-triplon state with large $\bar{N}$ do not
describe a spin-liquid state, in particular, a gapped Z$_2$ spin liquid:
In this case, the entanglement entropy obeys an area law with $c = -\gamma = - \ln 2$, 
where $\gamma$ is the topological entanglement entropy
\cite{review-balents}; such feature is found, e.g., for the
nearest-neighbor RVB state on the kagome and triangular lattices \cite{julia17};
again, for the many-triplon state, a transition from a VBS state to
a spin-liquid one with the triplon number $\bar{N}$ would be 
characterize by $c \rightarrow -\ln 2$.

In summary, we have studied a system of interacting triplons $b$, the
elementary excitations above a VBS ground state, described
by an effective boson model derived within the bond-operator
formalism. In particular, we chose the spin-$1/2$ $J_1$-$J_2$ AFM
Heisenberg model on a square lattice and focused on the possible
columnar VBS ground state. We found that a many-triplon state is stable, but
the lowest-energy one is constituted by a small number of triplons. 
Moreover, we also discussed the properties of many-triplon states
constituted by large number of triplons. 
In particular, the spin-spin correlation functions indicated that such
states might be characterized by a more homogeneous singlet pattern than
the columnar VBS ground state.
However, based on the mean-field results for the dimer order
parameters and the bipartite entanglement entropy, we concluded that
the many-triplon states with large triplon number $\bar{N}$ may not
describe a (gapped) spin-liquid phase.

It is important to emphasize that our conclusions about the nature of
the many-triplon states, in particular, the ones with large triplon
number $\bar{N}$, are related to a particular Heisenberg model and VBS
(reference) state. As mentioned in Sec.~\ref{sec:model}, it is not
clear, at the moment, whether the ground state of the $J_1$-$J_2$
model within the intermediate parameter region 
$0.4\, J_1 \lesssim J_2 \lesssim 0.6\, J_1$ is a VBS or a spin-liquid state. 
It would be interesting to apply the mean-field procedure discussed 
here to a Heisenberg model for which there are (strong) evidences for
a gapped spin-liquid phase, contrast the obtained results with the ones
derived here, and, in particular, to check whether the (possible) lowest-energy
many-triplon state is constituted by a large number of triplons.
A possible candidate is the spin-$1/2$ $J_1$-$J_2$ AFM Heisenberg
model on a {\sl triangular} lattice: 
although a more recent DMRG calculation pointed to a gapless spin-liquid
phase \cite{hu19}, previous DMRG simulations \cite{white15,hu15,saad16}
indicated that a gapped spin-liquid ground 
state may set in within the intermediate parameter region 
$0.07\, J_1 \lesssim J_2 \lesssim 0.15\, J_1$.

\acknowledgments

We thank A. O. Caldeira, E. Miranda, L. Leite, and M. Vojta for
helpful discussions and  
FAPESP, Project No.~2010/00479-6, for the partial financial support.

\appendix

\section{Effective boson model I in real space}
\label{ap:details-boso}

In this section, we quote the expression of the effective boson model
\eqref{h-effective} in terms of the singlet $s_i$ and triplet $t_{i \alpha}$ 
boson operators.

Substituting the (generalized) bond operator representation
\eqref{spin-bondop} into the Heisenberg model \eqref{ham-dimer}, it is
possible to show that the four terms of the Hamiltonian
\eqref{h-effective} read 
\begin{align}	
    \mathcal{H}_0 =& - \frac{3}{4} J_1 \sum_i s_i^\dagger s_i , 
\nonumber \\
    \mathcal{H}_2 =& \frac{J_1}{4} \sum_i t_{i \alpha  }^{\dagger} t_{i \alpha } 
             + \frac{1}{4} \sum_{i, \tau} \zeta_2(\tau) 
                 \left( s_i s_{i+\tau}^\dagger  t_{i \alpha}^ \dagger t_{i+\tau \alpha}  + {\rm H.c.}  \right.
\nonumber \\
                  & \left. + \; s_i^\dagger s_{i+\tau}^\dagger   t_{i\alpha} t_{i+\tau  \alpha} + {\rm H.c.} \right) ,
\nonumber \\ 
    \mathcal{H}_3 =&  \frac{i}{4}    \epsilon_{\alpha \beta \lambda} \sum_{i,\tau }\zeta_3(\tau)\left[  
                  \left( s_i^\dagger   t_{i \alpha} + t_{i \alpha}^\dagger s_i \right )
                           t_{i + \tau \beta}^\dagger  t_{i + \tau  \lambda}   \right.
\nonumber \\
                        &\left.   - \; (i \leftrightarrow i+\tau) \right] ,
\nonumber \\ 
   \mathcal{H}_4 =& -\frac{1}{4} \epsilon_{\alpha \beta \lambda} \: \epsilon_{\alpha \mu \nu}
            \sum_{i,\tau} \zeta_4(\tau)t_{i \beta}^\dagger  t_{i+\tau \mu}^\dagger   t_{i \lambda}  t_{i+\tau \nu},
\label{Heff-sites}
\end{align}
where the summation convention over repeated indices is implied and
the $\zeta_i(\tau)$ functions are defined as
\begin{eqnarray}
  \zeta_2(\tau) &=& 2(J_1 - J_2)\delta_{\tau, 2} - J_1\delta_{\tau, 1} 
                         - J_2\left( \delta_{\tau, 1+2} + \delta_{\tau,1-2} \right),
\nonumber \\
  \zeta_3(\tau) &=& J_1\delta_{\tau,1} + J_2\left( \delta_{\tau, 1+2} + \delta_{\tau,1-2} \right),
\nonumber \\
  \zeta_4(\tau) &=& 2(J_1 + J_2)\delta_{\tau, 2} + J_1\delta_{\tau, 1} 
                         + J_2\left( \delta_{\tau, 1+2} + \delta_{\tau,1-2} \right),
\nonumber 
\end{eqnarray}
with $\taub_n$ being the dimer nearest-neighbor vectors \eqref{tau-col}.

\section{Details: effective boson model II and the mean-field
            approximation for a system of $\bar{N}$ triplons }
\label{ap:details-mf}

In this section, we quote alternative expressions for the constant
$E_{40}$ [Eq.~\eqref{h40-b}],
the coefficients $A^{(4)}_\bk$ and $B^{(4)}_\bk$ [Eq.~\eqref{ab4k}] of
the quadratic term $\mathcal{H}_{24}$ [Eq.~\eqref{h24-b}],
the expressions of the coefficients of the mean-field
Hamiltonian \eqref{h44-b-mf}, in addition to provide some details of the
self-consistent problem related to the mean-field approximation
discussed in Sec.~\ref{sec:mf}. 

We start considering the constant $E_{40}$ and the quadratic term
$\mathcal{H}_{24}$.  
Since the bare quartic vertex \eqref{gammak} can be written as
\begin{equation}
   \gamma_{\bk - \bp} = -\frac{1}{2} \sum^4_{i=1} C_i \left[  f_i(\bk) f_i(\bp),
                                   + \bar{f}_i(\bk) \bar{f}_i(\bp) \right],
\label{gammak02}
\end{equation}
where the coefficients $C_i$ are defined as
\begin{equation}
 C_1 = J_1, \;\;\;\;\;
 C_2 = 2(J_1 + J_2), \;\;\;\;\;
 C_3 = C_4 = J_2,
\label{Ccoefs} 
\end{equation}
the functions $f_i(\bq)$ are given by 
\begin{eqnarray}
   f_1(\bp) &=& \cos(2p_x), \;\;\;\;\;
   f_2(\bp) = \cos p_y, 
\nonumber \\
&& \label{ffunctions} \\
   f_3(\bp) &=& \cos(2p_x + p_y), \;\;\;\;\;
   f_4(\bp) = \cos(2p_x - p_y), 
\nonumber
\end{eqnarray}
and the functions $\bar{f}_i(\bp) = f_i(\bp)$ with the replacement 
$\cos(x) \rightarrow \sin(x)$, 
it is interesting to define the following set of coefficients
\begin{equation}
   a_4(i,j) = \frac{1}{N'} \sum_\bp f_i(\bp) g_i(\bp),
\label{a4coefs}
\end{equation}
where $i,j = 1,2,3,4$ and the functions $g_i(\bp)$ are defined in
terms of the Bogoliubov coefficients \eqref{bogo-coef}, i.e,
\begin{widetext}
\begin{eqnarray}
g_1(\bp) &=& v^2_\bp = \frac{1}{2}\left(  -1 + \frac{A_\bp}{\omega_\bp}  \right),
\quad\quad\quad\quad
g_2(\bp) = u_\bp v_\bp = \frac{B_\bp}{2\omega_\bp}.
\label{gfunctions}
\end{eqnarray}
It is then possible to rewrite the constant $E_{40}$ [Eq.~\eqref{h40-b}] 
and the coefficients $A^{(4)}_\bk$ and $B^{(4)}_\bk$ [Eq.~\eqref{ab4k}] as 
\begin{eqnarray}
E_{40} &=& \frac{3}{4}N\sum^4_{i=1} C_i  \left[  a^2_4(i,1)  -  a^2_4(i,2)  \right],
\nonumber \\
&& \nonumber \\
 A^{(4)}_\bk   &=& \sum^4_{i=1} C_i \frac{f_i(\bk)}{\omega_\bk} 
                  \left[  a_4(i,1)A_\bk  -  a_4(i,2)B_\bk  \right],
\quad\quad\quad\quad
 B^{(4)}_\bk   = \sum^4_{i=1} C_i \frac{f_i(\bk)}{\omega_\bk} 
                  \left[  a_4(i,2)A_\bk  -  a_4(i,1)B_\bk  \right],
\label{ab4k02}
\end{eqnarray}
where the coefficients $A_\bk$ and $B_\bk$ are respectively given by
Eqs.~\eqref{ak} and \eqref{bk} and $\omega_\bk$ is the triplon energy
\eqref{omega-harmonic}.

Within a mean-field approximation, that takes into account both normal
$h_\bk$ and anomalous $\bar{h}_\bk$ expectation values \eqref{mf-par2},
one shows, after a long but straightforward algebra, that the quartic
term $\mathcal{H}_{44}$ [Eq.~\eqref{h44-b}] assumes the form \eqref{h44-b-mf},
where the constant $E_{44}$ and the coefficients $\Delta_{1,\bk}$ and
$\Delta_{2,\bk}$ read
\begin{eqnarray}
E_{44} &=& \frac{3}{N'}\sum_{\bk\,\bp} 
                  \gamma_{\bk - \bp} \left[  (2u_\bk v_\bk)(2 u_\bp v_\bp) 
                                     - (u^2_\bk + v^2_\bk)(u^2_\bp + v^2_\bp)   \right]\bar{h}_\bk \bar{h}_\bp
\nonumber \\
&+&  2\gamma_{\bk - \bp} \left[  (u^2_\bk + v^2_\bk) (2 u_\bp v_\bp) 
                    -   (2u_\bk v_\bk)(u^2_\bp + v^2_\bp) \right]\bar{h}_\bk h_\bp  
       + \gamma_{\bk - \bp} \left[ (u^2_\bk + v^2_\bk)(u^2_\bp +  v^2_\bp)  
                    -  (2u_\bk v_\bk)(2 u_\bp v_\bp)  \right] h_\bk h_\bp,
\nonumber \\
&& \label{e44}  \\
\Delta_{1,\bk} &=& -\frac{2}{N'}\sum_\bp 
      \gamma_{\bk - \bp} \left[  (2u_\bk v_\bk)(u^2_\bp + v^2_\bp) -  (u^2_\bk + v^2_\bk) (2 u_\bp v_\bp)  \right]\bar{h}_\bk
  + \gamma_{\bk - \bp} \left[ (u^2_\bk + v^2_\bk )(u^2_\bp + v^2_\bp) -   (2u_\bk v_\bk) (2 u_\bp v_\bp)  \right] h_\bk,
\nonumber \\
&& \label{delta1} \\
\Delta_{2,\bk} &=& -\frac{2}{N'}\sum_\bp 
      \gamma_{\bk - \bp} \left[ (2u_\bk v_\bk) (2 u_\bp v_\bp)  -  (u^2_\bk + v^2_\bk)  (u^2_\bp + v^2_\bp)  \right]\bar{h}_\bk 
   + \gamma_{\bk - \bp} \left[ (u^2_\bk + v^2_\bk ) (2 u_\bp v_\bp)  -   (2u_\bk v_\bk) (u^2_\bp + v^2_\bp)  \right] h_\bk,
\nonumber \\
&& \label{delta2}
\end{eqnarray}
with $\gamma_\bk$ being the bare quartic vertex \eqref{gammak},
$u_\bk$ and $v_\bk$ being the Bogoliubov coefficients \eqref{bogo-coef},
and $h_\bk$ and $\bar{h}_\bk$ being respectively the normal and
anomalous expectation values \eqref{mf-par2}.

Again, due to the property \eqref{gammak02},
it is useful to define the set of coefficients
\begin{equation}
    b_4(i,j) = \frac{1}{N'}\sum_\bp f_i(\bp) \bar{g}_j(\bp),
\end{equation}
where $i,j = 1,2,3,4$ and the functions $\bar{g}_i(\bp)$ are given in
terms of the Bogoliubov coefficients \eqref{bogo-coef} 
and the normal and anomalous expectation values
\eqref{mf-par2}:  
\begin{eqnarray}
\bar{g}_1(\bp) &=& ( u^2_\bp + v^2_\bp) \bar{h}(\bp)
                         = - \frac{A_\bp \bar{B}_\bp}{2 \omega_\bp \Omega_\bp},
\quad \quad\quad\quad 
\bar{g}_2(\bp) = 2u_\bp v_\bp \bar{h}(\bp)
                         = - \frac{B_\bp \bar{B}_\bp}{2 \omega_\bp \Omega_\bp},
\nonumber \\
&& \nonumber \\
\bar{g}_3(\bp) &=& ( u^2_\bp + v^2_\bp) h(\bp)
          = \frac{A_\bp}{2\omega_\bp}\left( -1 +  \frac{\bar{A}_\bp}{\Omega_\bp} \right),
\quad \quad 
\bar{g}_4(\bp) = 2u_\bp v_\bp  h(\bp)
          = \frac{B_\bp}{2\omega_\bp}\left( -1 +  \frac{\bar{A}_\bp}{\Omega_\bp} \right).
\end{eqnarray}
Then, it is easy to show that Eqs.~\eqref{e44}--\eqref{delta2} can be
rewritten as
\begin{eqnarray}
 E_{44} &=& -\frac{3}{4}N \sum_i C_i 
                  \left[  \left[ b_4(i,2) - b_4(i,3) \right]^2 - \left[ b_4(i,1) - b_4(i,4) \right]^2 
                  \right],   
\nonumber \\
&& \nonumber \\
 \Delta_{1,\bk}   &=& \sum_i C_i \frac{f_i(\bk)}{\omega_\bk} 
           \left[  \left( b_4(i,3)  -  b_4(i,2) \right)A_\bk  + \left( b_4(i,1)  -  b_4(i,4)  \right)B_\bk  \right],
\nonumber \\
&& \nonumber \\
  \Delta_{2,\bk}   &=& \sum_i C_i \frac{f_i(\bk)}{\omega_\bk} 
           \left[  \left( b_4(i,4)  -  b_4(i,1) \right)A_\bk  + \left( b_4(i,2)  -  b_4(i,3)  \right)B_\bk  \right],
\label{delta2-2}
\end{eqnarray}
where $C_i$ are the coefficients \eqref{Ccoefs}, $f_i(\bk)$ are the
functions \eqref{ffunctions}, $A_\bk$ and $B_\bk$ are respectively the coefficients
\eqref{ak} and \eqref{bk},  and $\omega_\bk$ is the triplon energy
\eqref{omega-harmonic}.

Due to the form of Eq.~\eqref{delta2-2}, we define a new set of
coefficients
\begin{eqnarray}
b_1(i) &\equiv& b_4(i,4) - b_4(i,1)
       = \frac{1}{2N'}\sum_\bp f_i(\bp)  \frac{1}{\omega_\bp \bar{\Omega}_\bp}
          \left[ B_\bp \left(  \bar{A}_\bp - \bar{\Omega}_\bp  \right)   +  A_\bp \bar{B}_\bp \right],
\label{defb1} \\
b_2(i) &\equiv& b_4(i,3) - b_4(i,2)
      = \frac{1}{2N'}\sum_\bp f_i(\bp)  \frac{1}{\omega_\bp \bar{\Omega}_\bp}
          \left[ A_\bp \left(  \bar{A}_\bp - \bar{\Omega}_\bp  \right)   +  B_\bp \bar{B}_\bp \right],
\label{defb2}
\end{eqnarray}
where $i = 1,2,3,4$, $\bar{A}_\bp$ and $\bar{B}_\bp$ are the
coefficients \eqref{abbark}, and $\bar{\Omega}_\bp$ is the energy
\eqref{omega-bcs} of the elementary excitations above the many-triplon
state \eqref{mf-wf}.
Finally, we recall Eq.~\eqref{nbar} that is related to the 
condition \eqref{constraint2}:
\begin{equation}
  \frac{\bar{N}}{N} = \frac{3}{4N'} \sum_\bk  
      \left( -1 + \frac{\bar{A}_\bp}{\Omega_\bp} \right). 
\label{nbar02}
\end{equation}
Equations \eqref{defb1}--\eqref{nbar02} define a self-consistent
problem that is numerically solved for a fixed value of the triplon
number $\bar{N}$ and the ratio $J_2/J_1$ of the exchange couplings. 
Such set of self-consistent equations allows us to calculate the
coefficients $b_1(i)$ and $b_2(i)$, with $i = 1,2,3,4$, and the
chemical potential $\bar{\mu}$, yielding the
energy \eqref{egs-bcs} and the excitation spectrum
\eqref{omega-bcs}.

\end{widetext}



\begin{thebibliography}{}
\bibitem{comment01}
For more details about the $J_1$-$J_2$ model on the square lattice, we
refer the reader, e.g., to the Introduction section of
Ref.~\cite{doretto14} and the references therein.
\end{thebibliography}
\end{document}